\documentclass[a4paper,5pt,intlimits,twopages]{iopart}
\expandafter\let\csname equation*\endcsname\relax
\expandafter\let\csname endequation*\endcsname\relax
\usepackage[tbtags]{amsmath}
\usepackage{amsthm,amssymb,dsfont,mathrsfs,bbm,xfrac,tabularx,booktabs}
\usepackage{cite}
\usepackage{mathtools}
\usepackage{comment}
\usepackage{multirow}
\usepackage{upgreek}
\usepackage[cal=esstix]{mathalfa}
\usepackage[usenames,dvipsnames]{xcolor}
\usepackage[caption=false]{subfig}
\usepackage[english]{babel}
\usepackage[export]{adjustbox}
\usepackage{geometry}
\usepackage{hyperref}
\hypersetup{pdfauthor={Parisi Perrupato Sicuro},pdftitle={Random matching on random regular graph},%
            colorlinks, linktocpage=true, pdfstartpage=3, pdfstartview=FitV,%
    breaklinks=true, pdfpagemode=UseNone, pageanchor=true, pdfpagemode=UseOutlines,%
    plainpages=false, bookmarksnumbered, bookmarksopen=true, bookmarksopenlevel=1,%
    hypertexnames=true, pdfhighlight=/O,%
    urlcolor=orange, linkcolor=blue, citecolor=red, 
        }
\usepackage{tikz}
\usetikzlibrary{patterns,decorations,arrows,shapes.geometric,arrows,mindmap,backgrounds,positioning,positioning,fit,backgrounds,positioning,arrows,matrix,calc}
\usetikzlibrary{shapes,backgrounds,mindmap,shadows,shapes.geometric,topaths,snakes,arrows,pgfplots.groupplots,fadings,calc,decorations.markings,positioning,chains,fit}
\tikzfading[name=fade l,left color=transparent!100,right color=transparent!0]
\tikzfading[name=fade r,right color=transparent!100,left color=transparent!0]
\tikzfading[name=fade d,bottom color=transparent!100,top color=transparent!0]
\tikzfading[name=fade u,top color=transparent!100,bottom color=transparent!0]
\usepackage{braket}
\usepackage{pgfplotstable}

\usepackage{lmodern}
\usepackage[T1]{fontenc}

\newcommand{\mV}{\mathcal{V}}

\newcommand{\GRRG}{{\mathbb{G}_{\text{RRG}}}}

\newcommand{\R}{\mathds{R}}

\newcommand{\N}{\mathds{N}}

\newcommand{\media}[1]{{\left\langle#1\right\rangle}}
\newcommand{\mediaE}[1]{{\mathbb E\left[#1\right]}}
\newcommand{\dd}[1]{\mathrm{d}#1}

\catcode`,\active

\catcode`\,12

\newdimen\nodeSize
\newdimen\nodeDist
\tikzset{position/.style args={#1:#2 from #3}{at=(#3.#1), anchor=#1+180, shift=(#1:#2)}}
\theoremstyle{plain}

\allowdisplaybreaks
\begin{document}
\title{Random-link matching problems on random regular graphs}

\author{Giorgio Parisi$^{1,2}$, Gianmarco Perrupato$^1$, Gabriele Sicuro$^1$}
\address{$^1$Dipartimento di Fisica, Sapienza Universit\`a di Roma, P.le A. Moro 2, I-00185, Rome, Italy}
\address{$^2$INFN -- Sezione di Roma1, CNR-IPCF UOS Roma Kerberos, P.le A. Moro 2, I-00185, Rome, Italy}
\eads{
\mailto{giorgio.parisi@roma1.infn.it},
\mailto{gianmarco.perrupato@uniroma1.it},
\mailto{gabriele.sicuro@for.unipi.it}
}
\begin{abstract}We study the random-link matching problem on random regular graphs, alongside with two relaxed versions of the problem, namely the fractional matching and the so-called ``loopy'' fractional matching. We estimated the asymptotic average optimal cost using the cavity method. Moreover, we also study the finite-size corrections due to rare topological structures appearing in the graph at large sizes. We estimate these contributions using the cavity approach, and we compare our results with the output of numerical simulations. The analysis also clarifies the meaning of the finite-size contributions appearing in the fully-connected version of the problem, that has been already analyzed in the literature.  \end{abstract}
\section{Introduction}
The matching problem is a classical combinatorial optimization problem that has been widely investigated by mathematicians and computer scientists \cite{lovasz2009matching}. The problem has also interesting relations with fundamental problems in physics: the correspondence between dimer covering problems and the Ising model on lattices, for example, has been first highlighted the Sixties \cite{Kasteleyn1967}. In the matching problem, the target is to find a dimer covering on a given graph, in such a way that no node remains unpaired. If the graph is weighted, then it is typically interesting to find the dimer covering of minimum or maximum ``cost'', defined as the sum of the weights of the selected edges. From the computational point of view, the matching problem is an ``easy problem'' belonging to the \textsf{P} complexity class, and polynomial algorithm are available for its solution \cite{papadimitriou1998combinatorial,Kuhn,Edmonds1965,Bayati2008}. It has been soon realized that solving a (weighted) matching problem is equivalent to finding the ground state of a system having many possible configuration (the possible matchings) of different ``energy''. This physical point of view has been widely exploited and applied to combinatorial optimization problems in general: an Hamiltonian is associated with the cost function that has to be minimized, and the ground state (corresponding to the optimal solution) is obtained considering the zero-temperature limit \cite{Kirkpatrick1983}.  

When the effectiveness of statistical physics ideas in the study of optimization problems became evident, the matching problem played the role of ``toy model'', due to its simplicity and relevance. In particular, the theory of disordered systems seems to be especially suitable for the study of the typical properties of optimization problems in presence of randomness \cite{mezard1987spin,Mezard2002,Krzakala2007,mezard2009information}. The exploration of this topic, that also inspired powerful numerical techniques for the algorithmic solution of these problems, started in the eighties, with the seminal works of Orland \cite{Orland1985} and M\'ezard and Parisi \cite{Mezard1985}, that tackled, as prototype problem, the random-link matching problem on the complete graph. They proved that the random-link matching problem can be fully investigated using the replica method and the cavity method in the limit of large number of vertices \cite{Mezard1986a}. Exact results have been obtained about the leading order cost and the finite-size corrections \cite{Mezard1987,Ratieville2002,Lucibello2018}, the fluctuation of the average optimal cost \cite{Malatesta2019}, and its embedding in the Euclidean space \cite{Caracciolo2014,Caracciolo2015s,Caracciolo2015}.

Despite its popularity in the fully-connected version, very few results are available on the random-link matching problem on sparse topologies. Zhou and Ou-Yang \cite{Zhou2003a} and subsequently Zdeborov\'a and M\'ezard \cite{Zdeborova2006} studied, using the cavity method, the number of maximum and perfect (unweighted) matchings on sparse graphs. Their results have been later rigorously proved by Bordenave, Lelarge and Salez \cite{Bordenave2013}. As far as we know, a study of the random weighted matching problem on the Bethe lattice is, instead, missing in the literature. 

In this paper, we investigate exactly this formulation of the matching problem using the cavity method, both at the leading order and at the level of finite-size corrections. This analysis will be, first of all, of methodological interest: indeed, we will check, for the first time to our knowledge, the effectiveness of the cavity method in the evaluation of the contribution to finite-size corrections of rare topological structures in a combinatorial optimization problem defined on a sparse graph. Such an approach, introduced in the study of spin glasses in Refs.~\cite{Montanari2005,Ferrari2013,Lucibello2014b}, allows us to go beyond the leading order using results (like the cavity fields distribution) obtained at the leading level. More recently, Coja-Oghlan and coworkers \cite{Coja-Oghlan2018} have rigorously grounded the cavity method results for a large class of models, characterizing in particular the finite-size corrections to the leading order free-energy and showing that they are indeed expressed as a sum on cycle contributions. Moreover, in the matching problem these topological contributions seem to be directly related to the finite-size corrections appearing in the fully-connected case, a fact suggested in Ref.~\cite{Lucibello2018} that will be investigated in the present paper, where this sparse-dense correspondence will be numerically verified.

The paper is organized as follows. In Section~\ref{sec:matching} we will introduce the random-link matching problem in full generality, and we will recall some of the known results about its average optimal cost and the corresponding finite-size corrections. We will describe the cavity approach to derive the asymptotic cost on a random regular graph, we will discuss how to compute, using the cavity method, the finite-size corrections, and we will compare our population dynamics results with the numerics. In Section~\ref{sec:fractional} we will introduce a useful auxiliary version of the problem, called random fractional matching problem, and we will study it by means of the same techniques. In Section~\ref{sec:loopy} we will further relax some constraints adopted in our problem to study the so-called ``loopy'' fractional matching, giving the cavity equations for its solution both at the leading level and at the subleading order. In Section~\ref{sec:zN} we will further comment on the $z\to N$ limit, where $z$ is the coordination of the Bethe lattice. Finally, in Section~\ref{sec:conclusioni} we will give our conclusions.

\section{The random-link matching problem}\label{sec:matching}
The random-link matching problem (MP) on a generic graph $G=(\mathcal V;\mathcal E)$ with vertex set $\mathcal V$ having cardinality $N=|\mathcal V|$, and edge set $\mathcal E\subseteq\mathcal V\times\mathcal V$ is defined as follows. For each edge $e\in\mathcal E$ of the graph we draw a random weight $w_e\in\R^+$ distributed with probability density function $\varrho(w)$, that we assume to have support over the positive real axis. We associate an occupancy variable $m_e\in\{0,1\}$ with each edge $e\in\mathcal E$, in such a way that $m_e=1$ if the edge is considered occupied, and $m_e=0$ otherwise. We search for a maximum cardinality matching that has minimum cost, i.e., for the set of values $M=\{m_e\}_e$ satisfying the constraints
\begin{equation}\label{constraints}
 m_e\in\{0,1\},\quad \sum_{e\to v}m_e\leq 1\ \forall v\in\mathcal V,
\end{equation}
that maximizes
\begin{equation}
 |M|\coloneqq\sum_e m_e
\end{equation}
and minimizes the cost:
\begin{equation}\label{costoEG}
 E_G[M]\coloneqq\frac{1}{|M|}\sum_e m_e w_e .
\end{equation}
In Eq.~\eqref{constraints} the sum runs over all edges incident to the vertex $v\in\mathcal V$. A matching is said to be perfect if $2|M|=N$: a perfect matching is not possible on a generic graph, unless the hypotheses of Tutte's theorem are fulfilled, see Ref.~\cite{lovasz2009matching}. Note that assuming an even $N$ does not guarantee the existence of a perfect matching. For example, the following graph
\begin{center}
\begin{tikzpicture}[scale=0.25]
\node[label=above:{},shape=circle, draw, inner sep=2pt] (0) at (0,0) [right] {};
\node[label=above:{},shape=circle,position=90:{2mm} from 0, draw, inner sep=2pt] (1a) {};
\node[label=above:{},shape=circle,position=90:{2mm} from 1a] (1) {};
\node[label=above:{},shape=circle,position=-30:{2mm} from 0, draw, inner sep=2pt] (2a) {};
\node[label=above:{},shape=circle,position=-30:{2mm} from 2a] (2) {};
\node[label=above:{},shape=circle,position=210:{2mm} from 0, draw, inner sep=2pt] (3a) {};
\node[label=above:{},shape=circle,position=210:{2mm} from 3a] (3) {};
\draw[thin] (0) -- (1a);
\draw[thin] (0) -- (2a);
\draw[thin] (0) -- (3a);
\node[label=left:{},shape=circle,position=-162:{2mm} from 1, draw, inner sep=2pt] (1b) {};
\node[label=above:{},shape=circle,position=-234:{2mm} from 1, draw, inner sep=2pt] (1c) {};
\node[label=above:{},shape=circle,position=-306:{2mm} from 1, draw, inner sep=2pt] (1d) {};
\node[label=above:{},shape=circle,position=-378:{2mm} from 1, draw, inner sep=2pt] (1e) {};
\draw[thin] (1b) -- (1a);\draw[thin] (1b) -- (1d);
\draw[thin] (1c) -- (1b);\draw[thin] (1c) -- (1e);
\draw[thin] (1c) -- (1d);
\draw[thin] (1d) -- (1e);
\draw[thin] (1a) -- (1e);
\node[label=above:{},shape=circle,position=78:{2mm} from 2, draw, inner sep=2pt] (2b) {};
\node[label=above:{},shape=circle,position=6:{2mm} from 2, draw, inner sep=2pt] (2c) {};
\node[label=above:{},shape=circle,position=-66:{2mm} from 2, draw, inner sep=2pt] (2d) {};
\node[label=above:{},shape=circle,position=-138:{2mm} from 2, draw, inner sep=2pt] (2e) {};
\draw[thin] (2b) -- (2a);\draw[thin] (2b) -- (2d);
\draw[thin] (2c) -- (2b);\draw[thin] (2c) -- (2e);
\draw[thin] (2c) -- (2d);
\draw[thin] (2d) -- (2e);
\draw[thin] (2a) -- (2e);
\node[label=above:{},shape=circle,position=-42:{2mm} from 3, draw, inner sep=2pt] (3b) {};
\node[label=above:{},shape=circle,position=-114:{2mm} from 3, draw, inner sep=2pt] (3c) {};
\node[label=above:{},shape=circle,position=-186:{2mm} from 3, draw, inner sep=2pt] (3d) {};
\node[label=above:{},shape=circle,position=-258:{2mm} from 3, draw, inner sep=2pt] (3e) {};
\draw[thin] (3b) -- (3a);\draw[thin] (3b) -- (3d);
\draw[thin] (3c) -- (3b);\draw[thin] (3c) -- (3e);
\draw[thin] (3c) -- (3d);
\draw[thin] (3d) -- (3e);
\draw[thin] (3a) -- (3e);
\end{tikzpicture}
\end{center}
is the only connected regular graph with $N=16$ and coordination equal to $3$ not allowing for a perfect matching. Given a graph $G$ and denoting by $\mathcal M(G)$ the set of all possible matching on $G$ of maximal cardinality, one might be interested in the average optimal cost (AOC), i.e. 
\begin{equation}
 E_G\coloneqq\mediaE{\min_{M\in \mathcal M(G)} E_G[M]},
\end{equation}
where the notation $\mediaE{\bullet}$ represents the average over the disorder. In 1985, M\'ezard and Parisi \cite{Mezard1985} considered 
the case $G=K_{N}$, complete graph with $|\mathcal V|=N$ vertices.
Using both the replica approach \cite{Mezard1985} and the cavity method \cite{Mezard1986a}, M\'ezard and Parisi were able to evaluate the AOC in this problem for $N\to+\infty$, finding that, if $\lim_{w\to 0}\varrho(w)=\sfrac{2}{N}$,
\begin{equation}
 E_{\infty}\coloneqq\lim_{\mathclap{N\to+\infty}}E_{K_{N}}=\frac{\zeta(2)}{2},
\end{equation}
where $\zeta(z)$ is the Riemann zeta function. The study of the finite-size corrections to this asymptotic value requires some effort and it has been performed in Refs.~\cite{Mezard1987,Ratieville2002,Lucibello2018}. In particular, adopting a weight distribution 
\begin{equation}
\varrho_N(w)=\frac{2}{N}\e^{-\frac{2w}{N}}\theta(w)\label{expn}
\end{equation}
with $\theta(\bullet)$ Heaviside function, we have
\begin{subequations}\label{costom}
\begin{equation}
 E_{K_{N}}=\frac{\zeta(2)}{2}+\frac{\zeta(2)}{2N}+\frac{2}{N}\sum_{\substack{p\geq 3\\\text{odd}}}\frac{I_{p}}{p}+o\left(\frac{1}{N}\right)=\frac{\zeta(2)}{2}-\frac{0.0674(1)}{N}+o\left(\frac{1}{N}\right),
\end{equation}
where, denoting by $x_{p+1}\equiv x_1$,
\begin{equation}\label{Ip}
 I_p\coloneqq\int_1^\infty\frac{\dd y}{2y}\prod_{i=1}^p\int_0^{1}\frac{\theta(1-x_ix_{i+1})\dd x_i}{x_i+y}=\frac{1}{p}+o\left(\frac{1}{p}\right).
\end{equation}
\end{subequations}
In Ref.~\cite{Lucibello2017} it has been shown that, using the weight distribution in Eq.~\eqref{expn}, the sub-subleading corrections scales in an anomalous way, i.e., as $N^{-\sfrac{3}{2}}$, and not as $N^{-2}$ as one might expect. Moreover, in Ref.~\cite{Lucibello2018} it has been observed that, if the constraints are relaxed and any value $0\leq m_e\leq 1$ is allowed, cycles appear in the optimal solution, and, at the same time, both the sum contribution in Eq.~\eqref{costom} and the anomalous correction disappear. If, finally, may allow a vertex to ``self-match'', the $\sfrac{1}{N}$ corrections disappear at all. These facts suggested that the $\sfrac{1}{N}$ corrections in Eq.~\eqref{costom} are due to cycles in the graph, and the sum in Eq.~\eqref{costom} runs over contributions corresponding to cycles of different length $p$. This hypothesis also justifies the presence of the $O\left(N^{-\sfrac{3}{2}}\right)$ correction that naturally appears imposing a cut-off to the sum in Eq.~\eqref{costom} to take into account that the possible length of cycles is bounded for finite $N$ \cite{Lucibello2017}. Surprisingly enough, if instead of $K_N$ the bipartite complete graph $K_{N,N}$ is considered, the form of the finite-size corrections strongly simplifies. In the bipartite complete graph, $2N$ vertices are divided into two disjoints sets having the same cardinality $N$, and an edge is present between two vertices if and only if they belong to different sets. Assuming the same weight distribution given in Eq.~\eqref{expn}, it is possible to prove that the average optimal cost in this case is given by  \cite{Parisi1998,Linusson2004,Nair2005}
\begin{equation}
    E_{K_{N,N}}=\sum_{i=1}^{N}\frac{1}{i^2}=\zeta(2)-\frac{1}{N}+o\left(\frac{1}{N}\right).
\end{equation}
Note that the expression holds for any value of $N$. The simplification, and the absence of anomalous correction terms, suggest once again that cycles are indeed relevant for finite-size correction (only even cycles appear in bipartite graphs).

The role of topological structures, e.g., paths, cycles and loops in the finite-size corrections clearly emerged in the study of disordered spin systems on random graphs \cite{Ferrari2013,Lucibello2014}. In these cases, the cavity approach made clear the correspondence inferred above, and the corrections were actually evaluated studying finite-size fluctuations in the graph topology with respect to the asymptotic infinite tree structure. For the same reasons, in the present paper, we will consider the random-link matching problem on the random regular graph ensemble $\GRRG(N,z)$, and we will study both the AOC in the thermodynamic limit and its finite-size corrections. We consider a uniform measure\footnote{The measure is uniform over all labeled graph \cite{Bollobas2001book}. Another possibility is to consider the uniform measure over all non-isomorphic regular graphs with $N$ vertices and coordination $z$. However it can be proved that the two definitions are equivalent in the thermodynamic limit \cite{McKay_1984}.} over all graphs with $N$ vertices whose coordination is exactly $z$. For fixed $z\geq 3$, an element of this ensemble $\GRRG(N,z)$ admits almost always a perfect matching if $zN$ is even \cite{Wormald}. Random regular graphs have cycles of typical length $\frac{\ln N}{\ln z}$ for $N\gg 1$ and in the $N\to+\infty$ limit at fixed $z$, the Caley tree of coordination $z$ is recovered. On the other hand, for $z=N-1$ the complete graph $K_N$ is obtained. Random regular graphs are therefore excellent candidates for the study of both the corrections due to cycles and finite-coordination effects.
On top of the random topology, we generate a random weight for each edge of the graph with distribution $\varrho(w)$, and we search for the optimal matching cost on the obtained weighted graph. The AOC is obtained averaging over both the weights and the graph topology realizations. In our problem, therefore, two sources of disorder appear, namely the topological disorder, due to the fact that the graph is randomly extracted from the $\GRRG(N,z)$ ensemble, and the link-weight disorder, due to the fact that the weights are i.i.d.~random variables distributed with probability density function $\varrho(w)$. The AOC density is evaluated averaging over both these sources of disorder,
\begin{equation}\label{costo}
E_z(N)\coloneqq \mediaE{\min_{M\in \mathcal M(G)} E_G[M]}=E_z+\frac{E_z^{(1)}}{N}+o\left(\frac{1}{N}\right),
\end{equation}
where $\lim_{N\to+\infty}E_z(N)=E_z$. In the following, we will assume an exponential probability density
\begin{equation}\label{distpesi}
 \varrho_z(w)\coloneqq\frac{2}{z}\e^{-\frac{2w}{z}}\theta(w).
\end{equation}

\subsection{The asymptotic cost}\label{Sec:asymptCost}
The cavity method is a natural candidate for the study of random optimization on sparse graphs \cite{Mezard2001,mezard2009information}. In the particular case of the MP, we start writing down a partition function associated with our problem on a graph $G=(\mathcal V;\mathcal E)$ extracted from $\GRRG(N,z)$, as
\begin{equation}\label{partizione}
Z(\beta)\coloneqq\sum_{\{m_e=0,1\}}\prod_{v\in\mV}\left[\mathbb I\left(\sum_{e\to v}m_e\leq 1\right)\e^{-\gamma\left(1-\sum_{e\to v}m_{e}\right)}\right] \e^{-\frac{\beta N}{2} E_G[M]}.
\end{equation}
The indicator function $\mathbb I(s)$ of a statement $s$ is such that $\mathbb I(s)=1$ if $s$ is true, and zero otherwise. Observe that we have introduced the Lagrange multiplier $\gamma$ to impose the fact that the optimal matching is perfect. The AOC can be therefore obtained by
\begin{equation}
E_z= - \lim_{N\to +\infty}\lim_{\beta\to+\infty}\lim_{\gamma\to+\infty}\frac{2\mediaE{\ln Z(\beta)}}{\beta N}.
\end{equation}

The variables of our problem are the quantities $m_e\in\{0,1\}$ defined on the edges of our weighted graph $G$. They have to satisfy the constraints in Eq.~\eqref{constraints}; moreover, a weight $\exp\left[-\gamma\left(1-\sum_{e\to v}m_e\right)\right]$ appears in the partition function in Eq.~\eqref{partizione} for each vertex of the graph $G$. It is well known that we can 
associate with the partition function a bipartite graph, called factor graph, involving two types of nodes, called variable nodes and functional nodes respectively. The variable nodes corresponds to the edges of $G$, and they are linked to functional nodes representing the constraint in Eq.~\eqref{constraints} and the reweighting that asymptotically impose the perfect matching:
\begin{equation*}
\begin{gathered}
\scalebox{0.6}{
\begin{tikzpicture}
\node[label=above:{$v$},fill=black,shape=rectangle,draw] (0) at (0,0) [right] {};
\node[label=above:{},shape=circle, position=0:{\nodeDist} from 0,draw,opacity=0.3] (1) {};
\node[label=above:{},shape=circle, position=120:{\nodeDist} from 0,draw,opacity=0.3] (2) {};
\node[label=above:{},shape=circle, position=240:{\nodeDist} from 0,draw,opacity=0.3] (3) {};
\draw[thin] (0) -- (1);
\draw[thin] (0) -- (2);
\draw[thin] (0) -- (3);
\end{tikzpicture}}
\end{gathered}
\equiv\mathbb{I}\left[\sum_{e\to v}m_e\leq 1\right]\e^{-\gamma\left(1-\sum_{e\to v}m_e\right)}.
\end{equation*}
Each functional node of this kind is linked to the variable nodes involved in the constraint, and uniquely corresponds to a vertex in the graph $G$. Moreover, each variable node is linked to an additional functional node corresponding to the $\e^{-\beta m_ew_e}$ reweighting \footnote{Here we are assuming that $\eta\coloneqq 2MN^{-1}$ is exactly equal to $1$. As mentioned before, almost surely $\eta=1$ for $N\to+\infty$ \cite{bollobas1986}. The possible presence of $O(\sfrac{1}{N})$ corrections in $\eta$ to this asymptotic limit might give additional $O(\sfrac{1}{N})$ finite-size corrections to the asymptotic cost. However, with the graph sizes considered in this paper, we never found connected regular graphs without perfect matching in around $10^7$ Monte Carlo steps, so we have assumed finite-$N$ corrections to $\eta=1$ to be of higher order.}
\begin{equation*}
\begin{gathered}
\scalebox{0.6}{
\begin{tikzpicture}
\node[label=below:{$e$},shape=circle,draw] (0) at (0,0) [right] {};
\node[label=above:{}, fill=gray!60, shape=rectangle, position=90:{0.5\nodeDist} from 0, draw] (A) {};
\node[label=above:{},shape=rectangle, position=180:{\nodeDist} from 0] (B) {};
\node[label=above:{},shape=rectangle, position=0:{\nodeDist} from 0] (C) {};
\draw[thin] (0) -- (A);
\draw [black,opacity=0.3,thick] (0) -- (B);
\draw [black,opacity=0.3,thick] (0) -- (C);
\end{tikzpicture}}
\end{gathered}
\equiv \e^{-\beta m_e w_e }.
\end{equation*}
We can finally construct our factor graph following the rules above starting from $G$. Pictorially,
\[\begin{gathered}
\begin{tikzpicture}
\node[fill=black,shape=circle,draw,inner sep=1.5pt] (0) at (0,0) {};
\node[label=above:{},fill=black,shape=circle,position=120:{0.35*\nodeDist} from 0,draw,inner sep=1.5pt] (a2) {};
\node[draw=none,position=30:{0.35*\nodeDist} from a2,inner sep=2pt] (a2a) {};
\node[draw=none,position=150:{0.35*\nodeDist} from a2,inner sep=2pt] (a2b) {};
\node[label=above:{},fill=black,shape=circle,position=240:{0.35*\nodeDist} from 0,draw,inner sep=1.5pt] (a3) {};
\node[draw=none,position=-30:{0.35*\nodeDist} from a3,inner sep=2pt] (a3a) {};
\node[draw=none,position=-150:{0.35*\nodeDist} from a3,inner sep=2pt] (a3b) {};
\draw[thin,gray] (a2a) -- (a2) -- (a2b);
\draw[thin,gray] (a3a) -- (a3) -- (a3b);
\draw[thin] (a3) -- (0) -- (a2);
\node[fill=black,shape=circle, position=0:10mm from 0,draw,inner sep=1.5pt] (i) {};
\node[label=above:{},fill=black,shape=circle,position=60:{0.35*\nodeDist} from i,draw,inner sep=1.5pt] (b4) {};
\node[label=above:{},fill=black,shape=circle,position=-60:{0.35*\nodeDist} from i,draw,inner sep=1.5pt] (b5) {};
\draw[thin] (b4) -- (i) -- (b5);
\node[draw=none,position=30:{0.4*\nodeDist} from b4,inner sep=2pt] (b4a) {};
\node[draw=none,position=150:{0.4*\nodeDist} from b4,inner sep=2pt] (b4b) {};
\node[draw=none,position=-30:{0.4*\nodeDist} from b5,inner sep=2pt] (b5a) {};
\node[draw=none,position=-150:{0.4*\nodeDist} from b5,inner sep=2pt] (b5b) {};
\draw[thin,gray] (b4a) -- (b4) -- (b4b);
\draw[thin,gray] (b5a) -- (b5) -- (b5b);
\draw[thin] (0) -- (i);
\end{tikzpicture}
\end{gathered}\Longrightarrow
\begin{gathered}
\begin{tikzpicture}
\node[fill=black,shape=rectangle,draw,inner sep=2pt] (0) at (0,0) {};
\node[shape=circle,draw,inner sep=2pt, position=0:5mm from 0] (e) {};
\node[label=above:{}, fill=gray!60, shape=rectangle, position=90:{0.2\nodeDist} from e, draw,inner sep=2pt] (A) {};
\node[label=above:{},fill=black,shape=rectangle,position=120:{0.7*\nodeDist} from 0,draw,inner sep=2pt] (a2) {};
\node[draw,shape=circle,position=45:{0.2*\nodeDist} from a2,inner sep=1pt] (a2a) {};
\node[draw,shape=circle,position=135:{0.2*\nodeDist} from a2,inner sep=1pt] (a2b) {};
\draw[thin] (a2b) -- (a2) -- (a2a);
\node[label=above:{},shape=circle, position=120:{0.2\nodeDist} from 0,draw,inner sep=2pt] (2) {};
\node[label=above:{}, fill=gray!60, shape=rectangle, position=30:{0.2\nodeDist} from 2, draw,inner sep=2pt] (A2) {};
\node[label=above:{},fill=black,shape=rectangle,position=240:{0.7*\nodeDist} from 0,draw,inner sep=2pt] (a3) {};
\node[draw,shape=circle,position=-45:{0.2*\nodeDist} from a3,inner sep=1pt] (a3a) {};
\node[draw,shape=circle,position=-135:{0.2*\nodeDist} from a3,inner sep=1pt] (a3b) {};
\draw[thin] (a3b) -- (a3) -- (a3a);
\node[label=above:{},shape=circle, position=240:{0.2\nodeDist} from 0,draw,inner sep=2pt] (3) {};
\node[label=above:{}, fill=gray!60, shape=rectangle, position=-30:{0.2\nodeDist} from 3, draw,inner sep=2pt] (A3) {};
\draw[thin] (e) -- (0) -- (2) -- (a2) -- (2) -- (A2);
\draw[thin] (0) -- (3) -- (a3) -- (3) -- (A3);
\node[fill=black, position=0:5mm from e,draw,inner sep=2pt] (i) {};
\node[label=above:{},fill=black,shape=rectangle,position=60:{0.7*\nodeDist} from i,draw,inner sep=2pt] (b4) {};
\node[draw,shape=circle,position=45:{0.2*\nodeDist} from b4,inner sep=1pt] (b4a) {};
\node[draw,shape=circle,position=135:{0.2*\nodeDist} from b4,inner sep=1pt] (b4b) {};
\draw[thin] (b4b) -- (b4) -- (b4a);
\node[label=above:{},shape=circle, position=60:{0.2\nodeDist} from i,draw,inner sep=2pt] (4) {};
\node[label=above:{}, fill=gray!60, shape=rectangle, position=150:{0.2\nodeDist} from 4, draw,inner sep=2pt] (B4) {};
\node[label=above:{},fill=black,shape=rectangle,position=-60:{0.7*\nodeDist} from i,draw,inner sep=2pt] (b5) {};
\node[draw,shape=circle,position=-45:{0.2*\nodeDist} from b5,inner sep=1pt] (b5a) {};
\node[draw,shape=circle,position=-135:{0.2*\nodeDist} from b5,inner sep=1pt] (b5b) {};
\draw[thin] (b5b) -- (b5) -- (b5a);
\node[label=above:{},shape=circle, position=-60:{0.2*\nodeDist} from i,draw,inner sep=2pt] (5) {};\node[label=above:{}, fill=gray!60, shape=rectangle, position=-150:{0.25\nodeDist} from 5, draw,inner sep=2pt] (B5) {};
\draw[thin] (i) -- (4) -- (b4) -- (4) -- (B4);
\draw[thin] (i) -- (e) -- (A);
\draw[thin] (i) -- (5) -- (b5) -- (5) -- (B5);
\end{tikzpicture}
\end{gathered}\]
For each edge $e=(i,j)$ in the graph $G$ we can define an ``incoming'' message $\mu^{i\to e}$ from the functional node $i$ to the variable node $e$  as the marginal probability distribution of the variable $m_e$ obtained removing any other edge with endpoint $e$ except $(i,e)$ in the factor graph. Pictorially,
\begin{equation*}
\begin{gathered}
\scalebox{0.8}{
\begin{tikzpicture}
\node[label=left:{$i$},fill=black,shape=rectangle,draw] (0) at (0,0) {};
\node[shape=circle,draw] (e) at (20mm,0) {$e$};
\node[label=above:{}, fill=gray!60, shape=rectangle, position=90:{0.5\nodeDist} from e, draw,opacity=0.3] (A) {};
\node[label=above:{},fill=black,shape=rectangle,position=120:{1.2*\nodeDist} from 0,draw,opacity=0.3] (a2) {};
\node[label=above:{},shape=circle, position=120:{0.3*\nodeDist} from 0,draw,opacity=0.3] (2) {};
\node[label=above:{}, fill=gray!60, shape=rectangle, position=30:{0.3\nodeDist} from 2, draw,opacity=0.3] (A2) {};
\node[label=above:{},fill=black,shape=rectangle,position=240:{1.2*\nodeDist} from 0,draw,opacity=0.3] (a3) {};
\node[label=above:{},shape=circle, position=240:{0.3*\nodeDist} from 0,draw,opacity=0.3] (3) {};
\node[label=above:{}, fill=gray!60, shape=rectangle, position=330:{0.3\nodeDist} from 3, draw,opacity=0.3] (A3) {};
\draw[thick,->,-latex] (0) -- node[fill=white] {$\mu^{i\to e}$}  (e);
\draw[thin,opacity=0.3] (0) -- (2) -- (a2) -- (2) -- (A2);
\draw[thin,opacity=0.3] (0) -- (3) -- (a3) -- (3) -- (A3);
\node[label=right:{$j$},fill=black, position=0:10mm from e,draw,opacity=0.3] (i) {};
\node[label=above:{},fill=black,shape=rectangle,position=60:{1.2*\nodeDist} from i,draw,opacity=0.3] (b4) {};
\node[label=above:{},shape=circle, position=60:{0.3*\nodeDist} from i,draw,opacity=0.3] (4) {};
\node[label=above:{}, fill=gray!60, shape=rectangle, position=150:{0.3\nodeDist} from 4, draw,opacity=0.3] (B4) {};
\node[label=above:{},fill=black,shape=rectangle,position=-60:{1.2*\nodeDist} from i,draw,opacity=0.3] (b5) {};
\node[label=above:{},shape=circle, position=-60:{0.3*\nodeDist} from i,draw,opacity=0.3] (5) {};\node[label=above:{}, fill=gray!60, shape=rectangle, position=-150:{0.3\nodeDist} from 5, draw,opacity=0.3] (B5) {};
\draw[thin,opacity=0.3] (i) -- (4) -- (b4) -- (4) -- (B4);
\draw[thin,opacity=0.3,dashed] (e) -- (A);
\draw[thin,opacity=0.3,dashed] (e) -- (i);
\draw[thin,opacity=0.3] (i) -- (5) -- (b5) -- (5) -- (B5);
\end{tikzpicture}}
\end{gathered}
\end{equation*}
For $N\gg 1$, the tree-like structure of $G$, inherited by the factor graph, allows us to write a recurrence equation for $\mu^{i\to e}$, namely
\begin{equation}
\mu^{i\to e}(m_e)\propto
\sum_{\substack{\{m_{\hat e}\}\\\hat e\neq e}}\mathbb I\left(\sum_{\substack{\hat e\to i
}}m_{\hat e}\leq 1\right)\exp\left[-\gamma\left(1-\sum_{\hat e\to i}m_{\hat e}\right)\right]\prod_{\substack{\hat e \to i \\\hat e=(k,i)\neq e}}\e^{-\beta m_{\hat e} w_{\hat e}}\mu^{k\to \hat e}(m_{\hat e}).\label{muie}
\end{equation}
In the expression above, we made the assumption that the marginal distribution corresponding to the edges $i\to \hat e $ such that $\hat e\neq e$ factorizes: this assumption is exact on trees \cite{mezard2009information}. We can introduce the cavity field $h^{i\to e}$ that parameterizes the message $\mu^{i\to e}(m_e)$ as
\begin{equation}\label{hbeta}
h^{i\to e}\coloneqq-\frac{1}{\beta}\ln\frac{\mu^{i\to e}(0)}{\mu^{i\to e}(1)}
=-\frac{1}{\beta}\ln\left[\e^{-\gamma}+\sum_{\mathclap{\substack{\hat e\to i\\\hat e=(k,i)\neq e}}}\e^{-\beta \left(w_{\hat e}-h^{k\to\hat e}\right)}\right].
\end{equation}
The marginal distribution of the variable $m_e$ is therefore parameterized as $\mu(m_e)\propto\exp\left[-\beta \left(w_{e}-h^{i\to e}-h^{j\to e}\right)m_e\right]$. Eq.~\eqref{hbeta} implies, for $\beta\to +\infty$ and $\gamma\to+\infty$,
\begin{equation}\label{cavita2}
h^{i\to e}=\min_{\mathclap{\substack{\hat e\to i\\\hat e=(k,i)\neq e}}}\left(w_e-h^{k\to\hat e}\right),
\end{equation}
where the minimum runs over $z-1$ edges incident to $j$. The node $i$ is then matched using the edge $e^*$ such that
\begin{equation}
 e^*=\arg\min_{\mathclap{\substack{\hat e\to i\\\hat e=(k,i)\neq e}}}\left(w_{\hat e}-h^{k\to \hat e}\right),
\end{equation}
or, equivalently, an edge $e=(i,j)$ is occupied if $w_e\leq h^{i\to e}+h^{j\to e}$. 
The equation above can be used to solve a specific instance of the problem by means of a message-passing algorithm \cite{Zdeborova2006,Bayati2008,mezard2009information}. Being interested in the average optimal cost, however, we will infer from Eq.~\eqref{cavita2} the distributional equation
\begin{equation}\label{cavita3}
 h\stackrel{d}{=}\min_{\mathclap{1\leq e\leq z-1}}\left(w_e-h_e\right)
\end{equation}
where the equality holds in the distributional sense, the $z-1$ quantities $\{w_e\}_e$ follow the distribution $\varrho_z(w)$ of the edge weights and the $z-1$ cavity fields $\{h_e\}_e$ have the same distribution $p_z(h)$ of $h$. By means of a population dynamics algorithm, and using Eq.~\eqref{cavita3}, the distribution of the cavity fields $p_z(h)$ can be numerically evaluated. The AOC for $N\to+\infty$ is then
\begin{equation}\label{costoEz}
 E_z=z\int_0^{+\infty}\dd w\,w\varrho_z(w)\!\iint\dd h_1\dd h_2\theta(h_1+h_2-w) p_z(h_1)p_z(h_2).
\end{equation}
Denoting by
\begin{equation}
    \Phi_z(x)\coloneqq \int_{x}^\infty p_z(h)\dd h,
\end{equation}
from Eq.~\eqref{cavita3} we can write
\begin{equation}\label{cavitaPhiz}
    \Phi_z(x)=\left(1-\mediaE{\Phi_z(w-x)}\right)^{z-1},
\end{equation}
where the average is taken over the variable $w$. In the Appendix we have solved this equation up to $o(\sfrac{1}{z})$ corrections to estimate the finite-$z$ corrections to the AOC, finding that, in the large-$z$ expansion of $E_z$, the $\sfrac{1}{z}$ term is identically zero,
\begin{equation}\label{EzMP}
 E_z=\frac{\zeta(2)}{2}+O\left(\frac{1}{z^2}\right),
\end{equation}
implying that
\begin{equation}
\lim_{z\to+\infty}\lim_{N\to+\infty}\left(E_{K_N}-E_z(N)\right)=0.
\end{equation}
This nontrivial sparse-dense correspondence is of the type discussed in Ref.~\cite{Korada2011}. The obtained AOC corresponds to the cost on the infinite tree, and finite-size corrections to $E_z$ are out of the reach of the approach above. The formula above holds for both the MP and the assignment problem that we will discuss below, due to the fact that the two problems are actually the same on trees, and they only differ at the level of finite-size corrections.

\begin{figure}
\centering
    \subfloat[AOC of the random-link matching problem on random regular graphs with $z\geq 3$ with exponential weight distribution given in Eq.~\eqref{distpesi}. Cavity results (black circles) are compared with the average optimal cost (red squares, shifted by $10^{-3}$ in the $x$ direction for the sake of clarity) obtained solving the problem on random regular graphs of size $N=10^5$. The continuous line is obtained from a cubic fit. \label{fig:cavitaleading}]{\includegraphics[height=0.42\textwidth]{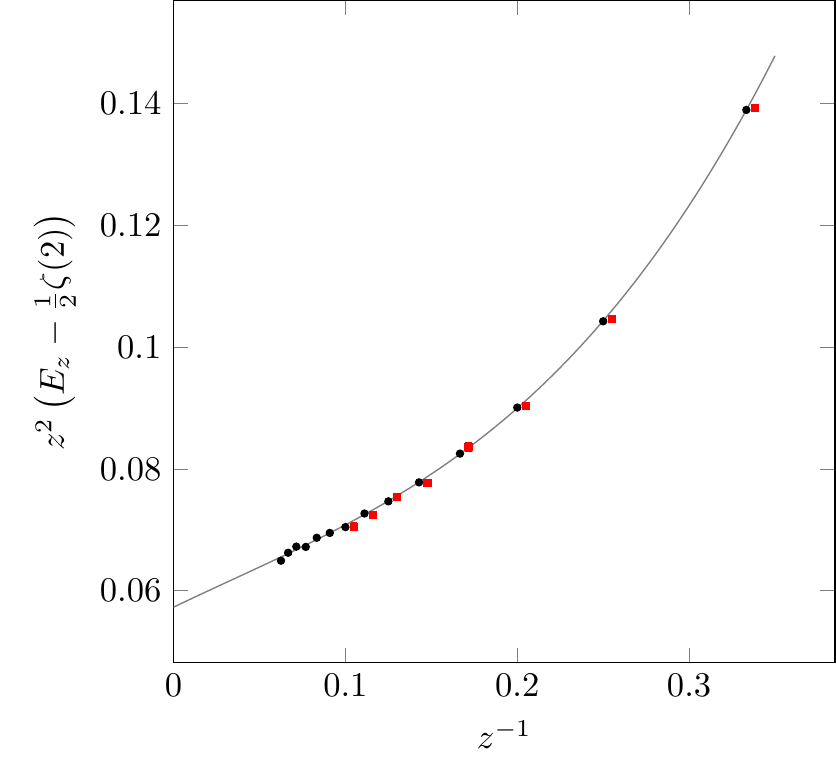}}\hfill
    \subfloat[Scaling of the finite-size correction for $z=3$ for the MP. See also Table ~\ref{tab:SummaryTable0}. The continuous line is a quadratic fit in $\sfrac{1}{\sqrt{N}}$.\label{fig:cavitaleading2}]{\includegraphics[height=0.42\textwidth]{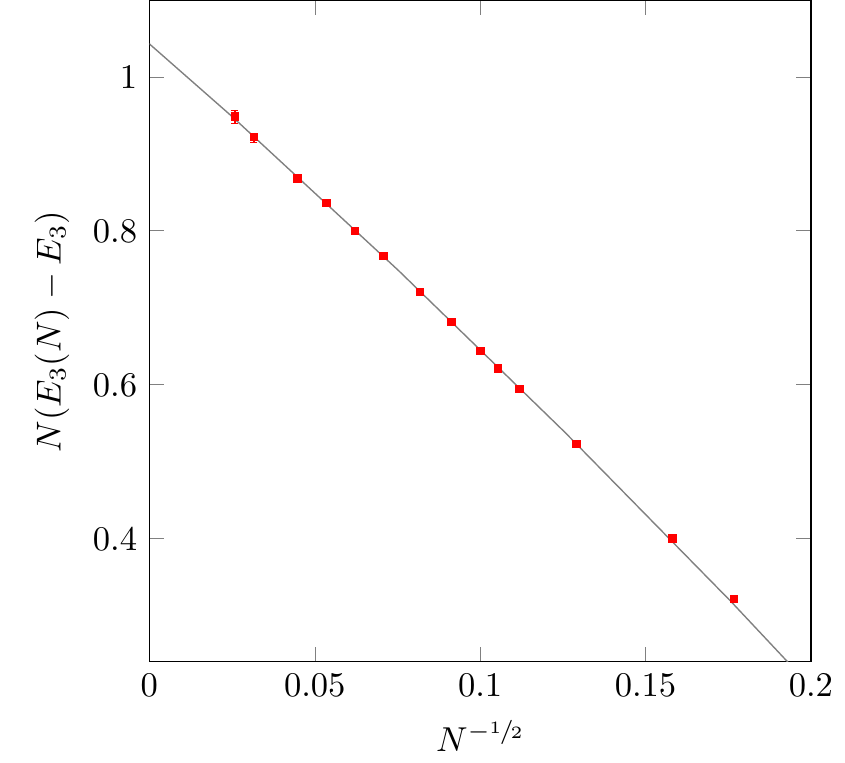}}
 \caption{
 Numerical and cavity results for the AOC in the random-link matching problem on random regular graphs. The cavity results have been obtained with a population of $10^7$ fields. For the numerical evaluation of the AOC of the random-link problems, we made use of the \texttt{Lemon} Graph Library \cite{lemon}. We mention here that, to improve the estimate of the AOC, we adopted a simple trick. Let us suppose that we want to estimate the average $\media{A}$ of some random scalar quantity $A$, and that we exactly know the average $\media{B}$ of a second scalar quantity $B$ that is positively correlated with $A$. Then, $\mathrm{Var}(A-B)=\mathrm{Var}(A)+\mathrm{Var}(B)-2\media{AB}+2\media{A}\media{B}$. It is possible therefore, depending on the choice of $B$, that $\mathrm{Var}(A)>\mathrm{Var}(A-B)$. In this case, evaluating $\media{A}$ as $\media{A-B}+\media{B}$ can be less noisy. We followed this approach for the AOC, choosing for $B$ the heuristic cost given selecting for each node the lowest weight incident edge. \label{fig:AOC}
 }
\end{figure}

\subsection{Finite-size corrections}\label{sec:correzioni}
As anticipated, the finite-size corrections are not directly accessible through the straightforward calculation given in the previous section; however, following Refs.~\cite{Ferrari2013,Lucibello2014}, we expect that they can be expressed in terms of contributions due to topological structures that appear in the graph with $O\left(\sfrac{1}{N}\right)$ density. In particular, in the case of random regular graphs we expect cycles to give a fundamental contribution\footnote{As noted in Ref.~\cite{Ferrari2013}, an additional contribution, due to chains in the graph, has to be considered in general, e.g.~in Erd\H{o}s--R\'enyi random graphs. In the random regular case, this contribution is absent due to the fact that there are no fluctuations in the vertex connectivity \cite{Lucibello2014}.}. Indeed, the number of cycles of length $\ell$ in an infinite random regular graph of coordination $z$ is a Poisson variable \cite{Wormland1981} with mean
\begin{equation}
    \bar n_\ell(z)=\frac{(z-1)^\ell}{2\ell}
\end{equation}
The fact that the density of cycles is infinitesimal for $N\to+\infty$ suggests that the contribution $\phi_\ell(z)$ of each cycle of length $\ell$ can be considered as independent from the others, due to the increasingly large distance of cycles for $N\gg 1$, and therefore we can evaluate it as the contribution of a single cycle embedded in an infinite tree. On the basis of these observations, we expect $E^{(1)}_z$ to have the form
\begin{equation}\label{sommacicligenerale}
 E_z^{(1)}=\upsilon_z+\sum_{\ell=3}^\infty \bar n_\ell(z)\phi_\ell(z),
\end{equation}
where $\upsilon_z$ is cycle-independent\footnote{The presence of an additional constant $\upsilon_z$ might depend on the choice of the weight distribution function $\varrho_z(w)$ and/or possible symmetries of the problem \cite{Ratieville2002,Lucibello2018}.}. The cycle contribution $\phi_\ell(z)$ can be therefore evaluated using the cavity approach as
\begin{equation}\label{phil1}
 \phi_\ell(z)=\lim_{\beta\to+\infty}\mediaE{-\frac{1}{\beta}\ln\frac{Z_\ell(\beta)}{Z_{\mathrm T}(\beta)}},
\end{equation}
where $Z_{\mathrm T}(\beta)$ is the partition function of the infinite tree with coordination $z$ and $Z_\ell(\beta)$ is the partition function for a $z$-regular graph that differs from the infinite tree because of the presence of a single cycle of length $\ell$. The average is taken on the weights appearing on the edges. It is important to note that both graphs can be obtained from the same object, i.e., a ``cavity'' tree with $2\ell+2$ ``cavity variable nodes'',
\[
\begin{gathered}
\scalebox{0.8}{
\begin{tikzpicture}

\node[shape=rectangle,fill=black, scale=0.6, draw] (0) at (0,0) [right] {};
\node[label=above:{},shape=circle,position=120:{0.45\nodeDist} from 0, scale=0.3,draw] (0A) {};
\node[label=above:{},shape=circle,position=240:{0.45\nodeDist} from 0, scale=0.3,draw] (0B) {};
\node[label=above:{},shape=circle,position=180:{0.45\nodeDist} from 0, scale=0.3,draw] (0C) {};
\draw[thin] (0) -- (0A);\draw[thin] (0) -- (0B);\draw[thin] (0) -- (0C);
\node[shape=rectangle,fill=black,position=0:{2\nodeDist} from 0, scale=0.6, draw] (1) {};
\node[label=above:{},shape=circle,position=70:{0.45\nodeDist} from 1, scale=0.3,draw] (1A) {};
\node[label=above:{},shape=circle,position=110:{0.45\nodeDist} from 1, scale=0.3,draw] (1B) {};
\draw[thin] (1B)-- (1) -- (1A);
\node[shape=rectangle,fill=black,position=0:{2\nodeDist} from 1, scale=0.6,draw] (2) {};
\node[label=above:{},shape=circle,position=70:{0.45\nodeDist} from 2, scale=0.3,draw] (2A) {};
\node[label=above:{},shape=circle,position=110:{0.45\nodeDist} from 2, scale=0.3,draw] (2B) {};
\draw[thin] (2B)-- (2) -- (2A);
\node[shape=rectangle,fill=black,position=0:{2\nodeDist} from 2, scale=0.6, draw] (3) {};
\node[label=above:{},shape=circle,position=70:{0.45\nodeDist} from 3, scale=0.3,draw] (3A) {};
\node[label=above:{},shape=circle,position=110:{0.45\nodeDist} from 3, scale=0.3,draw] (3B) {};
\draw[thin] (3B)-- (3) -- (3A);
\node[shape=rectangle,fill=black,position=0:{2\nodeDist} from 3, scale=0.6, draw] (4) {};
\node[label=above:{},shape=circle,position=60:{0.45\nodeDist} from 4, scale=0.3,draw] (4A) {};
\draw[thin] (4) -- (4A);
\node[label=above:{},shape=circle,position=-60:{0.45\nodeDist} from 4, scale=0.3,draw] (4B) {};
\node[label=above:{},shape=circle,position=0:{0.45\nodeDist} from 4, scale=0.3,draw] (4C) {};
\draw[thin] (4C) -- (4) -- (4B);
\node[shape=circle,scale=0.6,draw,label=below:{\footnotesize $m_1$}] (V0) at ($(0)!0.3!(1)$) {};
\node[label=above:{},shape=rectangle,fill=gray!60,position=90:{0.25\nodeDist} from V0, scale=0.6,draw] (V0f) {};
\draw[thin] (V0) -- (V0f);
\node[shape=circle,scale=0.6,draw,label=below:{\footnotesize $m_2$}] (V1o) at ($(0)!0.7!(1)$) {};
\node[shape=circle,scale=0.6,draw,label=below:{\footnotesize $m_3$}] (V1) at ($(1)!0.3!(2)$) {};
\node[label=above:{},shape=rectangle,fill=gray!60,position=90:{0.25\nodeDist} from V1, scale=0.6,draw] (V1f) {};
\draw[thin] (V1) -- (V1f);
\node[shape=circle,scale=0.6,draw,label=below:{\footnotesize $m_4$}] (V2o) at ($(1)!0.7!(2)$) {};
\node[shape=circle,scale=0.6,draw,label=below:{\footnotesize $m_5$}] (V2) at ($(2)!0.3!(3)$) {};
\node[label=above:{},shape=rectangle,fill=gray!60,position=90:{0.25\nodeDist} from V2, scale=0.6,draw] (V2f) {};
\draw[thin] (V2) -- (V2f);
\node[shape=circle,scale=0.6,draw,label=below:{\footnotesize $m_6$}] (V3o) at ($(2)!0.7!(3)$) {};
\node[shape=circle,scale=0.6,draw,label=below:{\footnotesize $m_7$}] (V3) at ($(3)!0.3!(4)$) {};
\node[label=above:{},shape=rectangle,fill=gray!60,position=90:{0.25\nodeDist} from V3, scale=0.6,draw] (V3f) {};
\draw[thin] (V3) -- (V3f);
\node[shape=circle,scale=0.6,draw,label=below:{\footnotesize $m_8$}] (V4) at ($(3)!0.7!(4)$) {};
\draw[thin] (0) -- (V0);
\draw[thin] (V1o) -- (1) -- (V1);
\draw[thin] (V2o) -- (2) -- (V2);
\draw[thin] (V3o) -- (3) -- (V3);
\draw[thin] (4) -- (V4);
\end{tikzpicture}
}
\end{gathered}
\]
Because of the tree-like structure, the partition function of such a cavity factor graph conditioned to a given configuration $(m_1,...,m_{2\ell+2})$ of the cavity variable nodes factorizes in $\ell+2$ contributions. We can also write the messages corresponding to each contribution. The marginal probability of these cavity variables can be written as
\begin{subequations}
\begin{align}\label{cavzz1}
\begin{gathered}
\scalebox{0.8}{
\begin{tikzpicture}
\node[shape=rectangle,fill=black, scale=0.6, draw,label=below:{\footnotesize $i$}] (0) at (0,0) [right] {};
\node[label=above:{},shape=circle,position=120:{0.45\nodeDist} from 0, scale=0.3,draw] (0A) {};
\node[label=above:{},shape=circle,position=240:{0.45\nodeDist} from 0, scale=0.3,draw] (0B) {};
\node[label=above:{},shape=circle,position=180:{0.45\nodeDist} from 0, scale=0.3,draw] (0C) {};
\draw[thin] (0) -- (0A);\draw[thin] (0) -- (0B);\draw[thin] (0) -- (0C);
\node[shape=circle,scale=0.6,draw,label=below:{\footnotesize $e$},position=0:{1.5\nodeDist} from 0] (V0) {};
\draw[thin] (0) -- (V0);
\end{tikzpicture}
}
\end{gathered}&=\frac{Z_1(m_e,\beta)}{Z_1(\beta)}=\mu^{i\to e}(m_e),\\
\label{cavzz2}
\begin{gathered}
\scalebox{0.8}{
\begin{tikzpicture}
\node[shape=rectangle,fill=black, scale=0.6, draw,label=below:{\footnotesize $i$}] (0) at (0,0) [right] {};
\node[label=above:{},shape=circle,position=120:{0.45\nodeDist} from 0, scale=0.3,draw] (0A) {};
\node[label=above:{},shape=circle,position=240:{0.45\nodeDist} from 0, scale=0.3,draw] (0B) {};
\node[label=above:{},shape=circle,position=180:{0.45\nodeDist} from 0, scale=0.3,draw] (0C) {};
\draw[thin] (0) -- (0A);\draw[thin] (0) -- (0B);\draw[thin] (0) -- (0C);
\node[shape=circle,scale=0.6,draw,label=below:{\footnotesize $e$},position=0:{1.5\nodeDist} from 0] (V0) {};
\node[label=above:{},shape=rectangle,fill=gray!60,position=90:{0.25\nodeDist} from V0, scale=0.6,draw] (V0f) {};
\draw[thin] (0) -- (V0) -- (V0f);
\end{tikzpicture}
}
\end{gathered}&=\frac{Z'_{1}(m_e,\beta)}{Z'_{1}(\beta)}\propto \e^{-\beta m_ew_e}\mu^{i\to e}(m_e),
\\ \label{cavzz3}   \begin{gathered}
\scalebox{0.8}{
\begin{tikzpicture}
\node[shape=rectangle,fill=black, scale=0.6, draw,label=below:{\footnotesize $i$}] at (0,0) (3) {};
\node[label=above:{},shape=circle,position=70:{0.45\nodeDist} from 3, scale=0.3,draw] (3A) {};
\node[label=above:{},shape=circle,position=110:{0.45\nodeDist} from 3, scale=0.3,draw] (3B) {};
\draw[thin] (3B)-- (3) -- (3A);
\node[shape=circle,scale=0.6,draw,label=below:{\footnotesize $e$},position=180:{\nodeDist} from 3] (V3o) {};
\node[shape=circle,scale=0.6,draw,label=below:{\footnotesize $\tilde e$},position=0:{\nodeDist} from 3] (V3) {};
\node[label=above:{},shape=rectangle,fill=gray!60,position=90:{0.25\nodeDist} from V3, scale=0.6,draw] (V3f) {};
\draw[thin] (V3o) -- (3) -- (V3) -- (V3f);
\end{tikzpicture}
}
\end{gathered}&=\frac{Z_2(m_{e},m_{\tilde e},\beta)}{Z_2(\beta)}\equiv\hat \mu^{i\to e,\tilde e}(m_e,m_{\tilde e}).
\end{align}
\label{cavZZZ}
\end{subequations}
Here $Z_2(\beta)$ is the partition function of the graph containing two cavity variables, whereas $Z_2(m_1,m_2,\beta)$ is the partition function of the same graph obtained constraining the cavity variables to have value $m_1$ and $m_2$. Similar notation holds for $Z_1$ and $Z_1'$. The marginals defined in Eq~\eqref{cavzz1} and Eq.~\eqref{cavzz2} are exactly the messages in Eq.~\eqref{muie}. The marginal in Eq.~\eqref{cavzz3} satisfies instead the equation
\begin{equation}\label{hhat1}
\hat \mu^{i\to e,\tilde e}(m_e,m_{\tilde e})\propto\e^{-\beta m_{\tilde e}w_{\tilde e}}\sum_{\mathclap{\{m_{\hat e}\}_{\hat e\neq e,\tilde e}}}\mathbb I\left(\sum_{\hat e\to i}m_{\hat e}=1\right)\prod_{\mathclap{\substack{\hat e\to i\\\hat e=(k,i)\neq e,\tilde e}}}\e^{-\beta m_{\hat e}w_{\hat e}}\mu^{k\to\hat e}(m_{\hat e}).
\end{equation}
This marginal is such that $\hat \mu^{i\to e,\tilde e}(1,1)=0$. The normalization constraint leaves us two parameters to parametrize the distribution; moreover
\begin{equation}
-\frac{1}{\beta}\ln\frac{\hat \mu^{i\to e,\tilde e}(0,1)}{\hat \mu^{i\to e,\tilde e}(1,0)}=w_{\tilde e}.
\end{equation}
We can therefore parametrize the distribution with one parameter only, writing
\begin{equation}\label{hhat2}
\hat \mu^{i\to e,\tilde e}(m_e,m_{\tilde e})\propto \e^{-\beta m_{\tilde e}w_{\tilde e}+\beta \hat h^{i\to e,\tilde e}(1-m_{e}m_{\tilde e})}\mathbb I(m_em_{\tilde e}\neq 1). 
\end{equation}
The partition function of the entire object is therefore $Z_\text{cav}(\beta)=Z_1(\beta)Z_1'(\beta)Z_2^\ell(\beta)$. On the other hand, from the cavity graph we can reconstruct original tree identifying $m_{2i-1}$ and $m_{2i}$ for $i=1,\dots,\ell+1$, obtaining the ratio between the partition function of the tree $Z_\text{T}(\beta)$ and  $Z_\text{cav}(\beta)$,
\begin{equation}
\begin{gathered}
\scalebox{0.8}{
\begin{tikzpicture}
\node[shape=rectangle,fill=black, scale=0.6, draw] (0) at (0,0) [right] {};
\node[label=above:{},shape=circle,position=120:{0.45\nodeDist} from 0, scale=0.3,draw] (0A) {};
\node[label=above:{},shape=circle,position=240:{0.45\nodeDist} from 0, scale=0.3,draw] (0B) {};
\node[label=above:{},shape=circle,position=180:{0.45\nodeDist} from 0, scale=0.3,draw] (0C) {};
\draw[thin] (0) -- (0A);\draw[thin] (0) -- (0B);\draw[thin] (0) -- (0C);
\node[shape=rectangle,fill=black,position=0:{\nodeDist} from 0, scale=0.6, draw] (1) {};
\node[label=above:{},shape=circle,position=90:{0.45\nodeDist} from 1, scale=0.3,draw] (1A) {};
\node[label=above:{},shape=circle,position=-90:{0.45\nodeDist} from 1, scale=0.3,draw] (1B) {};
\draw[thin] (1B)-- (1) -- (1A);
\node[shape=rectangle,fill=black,position=0:{\nodeDist} from 1, scale=0.6,draw] (2) {};
\node[label=above:{},shape=circle,position=90:{0.45\nodeDist} from 2, scale=0.3,draw] (2A) {};
\node[label=above:{},shape=circle,position=-90:{0.45\nodeDist} from 2, scale=0.3,draw] (2B) {};
\draw[thin] (2B)-- (2) -- (2A);
\node[shape=rectangle,fill=black,position=0:{\nodeDist} from 2, scale=0.6, draw] (3) {};
\node[label=above:{},shape=circle,position=90:{0.45\nodeDist} from 3, scale=0.3,draw] (3A) {};
\node[label=above:{},shape=circle,position=-90:{0.45\nodeDist} from 3, scale=0.3,draw] (3B) {};
\draw[thin] (3B)-- (3) -- (3A);
\node[shape=rectangle,fill=black,position=0:{\nodeDist} from 3, scale=0.6, draw] (4) {};
\node[label=above:{},shape=circle,position=60:{0.45\nodeDist} from 4, scale=0.3,draw] (4A) {};
\draw[thin] (4) -- (4A);
\node[label=above:{},shape=circle,position=-60:{0.45\nodeDist} from 4, scale=0.3,draw] (4B) {};
\node[label=above:{},shape=circle,position=0:{0.45\nodeDist} from 4, scale=0.3,draw] (4C) {};
\draw[thin] (4C) -- (4) -- (4B);
\node[shape=circle,scale=0.6,draw] (V0) at ($(0)!0.5!(1)$) {};
\node[label=above:{},shape=rectangle,fill=gray!60,position=90:{0.25\nodeDist} from V0, scale=0.6,draw] (V0f) {};
\draw[thin] (V0) -- (V0f);
\node[shape=circle,scale=0.6,draw] (V1) at ($(1)!0.5!(2)$) {};
\node[label=above:{},shape=rectangle,fill=gray!60,position=90:{0.25\nodeDist} from V1, scale=0.6,draw] (V1f) {};
\draw[thin] (V1) -- (V1f);
\node[shape=circle,scale=0.6,draw] (V2) at ($(2)!0.5!(3)$) {};
\node[label=above:{},shape=rectangle,fill=gray!60,position=90:{0.25\nodeDist} from V2, scale=0.6,draw] (V2f) {};
\draw[thin] (V2) -- (V2f);
\node[shape=circle,scale=0.6,draw] (V3) at ($(3)!0.5!(4)$) {};
\node[label=above:{},shape=rectangle,fill=gray!60,position=90:{0.25\nodeDist} from V3, scale=0.6,draw] (V3f) {};
\draw[thin] (V3) -- (V3f);
\draw[thin] (0) -- (V0) -- (1) -- (V1) --(2) -- (V2) --(3) -- (V3) --(4);
\end{tikzpicture}
}
\end{gathered}=\frac{\sum_{\{m_i\}_{i=1}^{\ell+1}}Z_1(m_1,\beta)Z_1'(m_{\ell+1},\beta)\prod_{i=2}^{\ell+1} Z_2(m_{i},m_{i+1},\beta)}{Z_\text{cav}(\beta)}\equiv\frac{Z_\text{T}(\beta)}{Z_\text{cav}(\beta)}.
\end{equation}
Similarly, in order to get $Z_{\ell}(\beta)Z^{-1}_\text{cav}(\beta)$, we can construct from the cavity graph a topology containing a loop
\begin{equation}\left[
\begin{gathered}
\scalebox{0.8}{
\begin{tikzpicture}
\node[shape=rectangle,fill=black, scale=0.6 ,draw] (A) at (0,0) [right] {};
\node[shape=rectangle,fill=black,position=0:{\nodeDist} from A, scale=0.6, draw,draw] (B) {};
\node[shape=rectangle,fill=black,position=60:{\nodeDist} from A, scale=0.6, draw,draw] (C) {};
\node[shape=circle,scale=0.6,draw] (V1) at ($(A)!0.5!(B)$) {};
\node[label=above:{},shape=rectangle,fill=gray!60,position=-90:{0.25\nodeDist} from V1, scale=0.6,draw] (V1f) {};
\draw[thin] (V1) -- (V1f);
\node[shape=circle,scale=0.6,draw] (V2) at ($(B)!0.5!(C)$) {};
\node[label=above:{},shape=rectangle,fill=gray!60,position=30:{0.25\nodeDist} from V2, scale=0.6,draw] (V2f) {};
\draw[thin] (V2) -- (V2f);
\node[shape=circle,scale=0.6,draw] (V3) at ($(C)!0.5!(A)$) {};
\node[label=above:{},shape=rectangle,fill=gray!60,position=150:{0.25\nodeDist} from V3, scale=0.6,draw] (V3f) {};
\draw[thin] (V3) -- (V3f);
\node[label=above:{},shape=circle,position=180:{0.45\nodeDist} from A, scale=0.3,draw] (1A) {};
\node[label=above:{},shape=circle,position=240:{0.45\nodeDist} from A, scale=0.3,draw] (2A) {};
\draw[thin] (2A) -- (A) -- (1A);
\node[label=above:{},shape=circle,position=-60:{0.45\nodeDist} from B, scale=0.3,draw] (1B) {};
\node[label=above:{},shape=circle,position=0:{0.45\nodeDist} from B, scale=0.3,draw] (2B) {};
\draw[thin] (2B) -- (B) -- (1B);
\node[label=above:{},shape=circle,position=60:{0.45\nodeDist} from C, scale=0.3,draw] (1C) {};
\node[label=above:{},shape=circle,position=120:{0.45\nodeDist} from C, scale=0.3,draw] (2C) {};
\draw[thin] (2C) -- (C) -- (1C);
\draw[thin] (A) -- (V1) -- (B) -- (V2) --(C) -- (V3) --(A);
\end{tikzpicture}
}
\end{gathered}
\quad
\begin{gathered}
\scalebox{0.8}{
\begin{tikzpicture}
\node[shape=rectangle,fill=black, scale=0.6 ,draw] (A) at (0,0) [right] {};
\node[label=above:{},shape=circle,position=120:{0.45\nodeDist} from A, scale=0.3,draw] (1A) {};
\draw[thin] (A) -- (1A);
\node[label=above:{},shape=circle,position=240:{0.45\nodeDist} from A, scale=0.3,draw] (2A) {};
\node[label=above:{},shape=circle,position=180:{0.45\nodeDist} from A, scale=0.3,draw] (3A) {};
\draw[thin] (3A) -- (A) -- (2A);
\node[shape=rectangle,fill=black,position=0:{\nodeDist} from A, scale=0.6,draw] (B) {};
\node[label=above:{},shape=circle,position=60:{0.45\nodeDist} from B, scale=0.3,draw] (1B) {};
\draw[thin] (B) -- (1B);
\node[label=above:{},shape=circle,position=-60:{0.45\nodeDist} from B, scale=0.3,draw] (2B) {};
\node[label=above:{},shape=circle,position=0:{0.45\nodeDist} from B, scale=0.3,draw] (3B) {};
\draw[thin] (3B) -- (B) -- (2B);
\node[shape=circle,scale=0.6,draw] (V) at ($(A)!0.5!(B)$) {};
\node[label=above:{},shape=rectangle,fill=gray!60,position=90:{0.25\nodeDist} from V, scale=0.6,draw] (Vf) {};
\draw[thin] (V) -- (Vf);
\draw[thin] (B) -- (V) -- (A);
\end{tikzpicture}
}
\end{gathered}\right]=\frac{\sum_m Z_1(m,\beta)Z_1'(m,\beta)\sum_{\{m_i\}_{i=1}^\ell}\prod_{i=1}^\ell Z_2(m_i,m_{i+1},\beta)}{Z_\text{cav}(\beta)}\equiv\frac{Z_\ell(\beta)}{Z_\text{cav}(\beta)}
\end{equation}
(where in the product $m_{\ell+1}\equiv m_1$). Using Eqs.~\eqref{cavZZZ}, it can be easily shown that the specific contribution of the cycle can be expressed as a trace of products of random matrices as follows:
\begin{equation}
Z^c_\ell(\beta)\coloneqq\sum_{\{m_i\}_{i=1}^\ell}\prod_{i=1}^\ell \frac{Z_2(m_i,m_{i+1},\beta)}{Z_2(\beta)}=\tr{\prod_{i=1}^\ell \mathbf T_i},\qquad 
 \mathbf T_{i}
 \coloneqq \begin{pmatrix}
 \e^{\beta \hat h_i}&1\\\e^{-\beta w_i}&0
 \end{pmatrix},
\end{equation}
where $\hat h_i$ are the cavity fields entering the cycle. Note that this structure is reminiscent of Eq.~\eqref{Ip}. The study of the spectral properties of the transfer matrix $\mathbf T$, performed using the replica trick, can provide an estimate of the large $\ell$ behavior $\phi_\ell$ \cite{Ferrari2013,Lucibello2014}. Due to the fact that in our problem the variables live on the edges, and not on the vertices of the graph, from the factor graph point of view, the cycle is connected to the rest of the graph through its functional nodes (and not through its variable nodes). More importantly, the messages entering the cycle are determined by the infinite tree-like structure. Similarly, a chain contribution can be written as
\begin{equation}
Z^a_\ell(\beta)\coloneqq \sum_{\{m_i\}_{i=1}^{\ell}}\frac{Z_1(m_1,\beta)}{Z_1(\beta)}\frac{Z_1'(m_{\ell},\beta)}{Z_1'(\beta)}\prod_{i=2}^{\ell-1} \frac{Z_2(m_{i},m_{i+1},\beta)}{Z_2(\beta)}=\mathbf u_1\cdot\left(\prod_{i=2}^{\ell-1}\mathbf T_i\right)\cdot\mathbf u_\ell,
\end{equation}
where $\mathbf u_i=(\e^{\beta h_i},\e^{-\beta w_i})$. 

In the $\beta\to+\infty$ limit, the equations strongly simplify. First observe that, from Eq.~\eqref{hhat1} and Eq.~\eqref{hhat2} we obtain
\begin{equation}
\hat h^{i\to e,\tilde e}=\min_{\substack{\hat e\to i\\\hat e\neq e,\tilde e}}(w_{\hat e}-h^{k\to i}),
\end{equation}
The field $\hat h$ corresponds to the outgoing cavity field from a functional node having $z-2$ neighbours. In the $\beta\to\infty$ limit, the fields $\hat h$ satisfy the distributional equation 
\begin{equation}
    \hat h\stackrel{d}{=}\min_{\mathclap{1\leq e\leq z-2}}\left(w_e-h_e\right).
\end{equation}
We can define the chain cost $\phi^a_\ell$ and the cycle cost $\phi^c_\ell$ as 
\begin{equation}
\phi^a_\ell(z)\coloneqq-\lim_{\beta\to +\infty}\frac{\mediaE{\ln Z_\ell^a(\beta)}}{\beta},\qquad \phi^c_\ell(z)\coloneqq -\lim_{\beta\to+\infty}\frac{\mediaE{\ln Z_\ell^c(\beta)}}{\beta},
\end{equation}
and write an equation that is similar to the one appearing in the cycle expansion of finite-size corrections of spin-glass systems on random graphs \cite{Ferrari2013,Lucibello2014}
\begin{equation}
    \phi_\ell(z)=\phi^c_\ell(z)-(\phi^a_{\ell+2}(z)-\phi^a_{2}(z)).
\end{equation}
The $\beta\to+\infty$ limit can be numerically evaluated considering, for each cycle or chain of length $\ell$, all possible ways of occupying it, and averaging over the minimum cost configurations: each internal edge contributes with its weight, each external edge entering the matching gives (minus) the corresponding cavity field. For example, for $\ell=3$,
\[\phi_3^c(z)=\mathbb E\left[\min\left\{
\underbrace{\parbox[c]{30pt}{\begin{tikzpicture}[auto,
  nodeDecorate/.style={shape=circle,inner sep=1pt,draw,thick},%
  lineDecorate/.style={-,thick}]
\draw (0:0.25) node[nodeDecorate] (v0) {\tiny $1$};
\draw (120:0.25) node[nodeDecorate] (v120) {\tiny $2$};
\draw (240:0.25) node[nodeDecorate] (v240) {\tiny $3$};
\foreach \x in {0,120,...,359}{
\draw (\x:0.55) node[nodeDecorate,fill=red!50,inner sep=2pt] (e\x) {};}
\draw[lineDecorate,opacity=0.3] (v0) -- (v120);
\draw[lineDecorate,opacity=0.3] (v240) -- (v0);
\draw[lineDecorate,opacity=0.3] (v240) -- (v120);
\draw[lineDecorate,red] (v120) -- (e120);
\draw[lineDecorate,red] (v240) -- (e240);
\draw[lineDecorate,red] (v0) -- (e0);
\end{tikzpicture}}}_{-\hat h_1-\hat h_2-\hat h_3},
\underbrace{
\parbox[c]{30pt}{\begin{tikzpicture}[auto,
  nodeDecorate/.style={shape=circle,inner sep=1pt,draw,thick},%
  lineDecorate/.style={-,thick}]
\draw (0:0.25) node[nodeDecorate] (v0) {\tiny $1$};
\draw (120:0.25) node[nodeDecorate] (v120) {\tiny $2$};
\draw (240:0.25) node[nodeDecorate] (v240) {\tiny $3$};
\foreach \x in {0,120,...,359}{
\draw (\x:0.55) node[nodeDecorate,fill=gray,inner sep=2pt,opacity=0.3] (e\x) {};}
\draw (e120) node[nodeDecorate,fill=red!50,inner sep=2pt] {};
\draw[lineDecorate,opacity=0.3] (v120) -- (v240);
\draw[lineDecorate,opacity=0.3] (v120) -- (v0);
\draw[lineDecorate,red] (v240) -- (v0);
\draw[lineDecorate,opacity=0.3] (v0) -- (e0);
\draw[lineDecorate,opacity=0.3] (v240) -- (e240);
\draw[lineDecorate,red] (v120) -- (e120);
\end{tikzpicture}}}_{w_{13}-\hat h_2},
\underbrace{
\parbox[c]{30pt}{\begin{tikzpicture}[auto,
  nodeDecorate/.style={shape=circle,inner sep=1pt,draw,thick},%
  lineDecorate/.style={-,thick}]
\draw (0:0.25) node[nodeDecorate] (v0) {\tiny $1$};
\draw (120:0.25) node[nodeDecorate] (v120) {\tiny $2$};
\draw (240:0.25) node[nodeDecorate] (v240) {\tiny $3$};
\foreach \x in {0,120,...,359}{
\draw (\x:0.55) node[nodeDecorate,fill=gray,inner sep=2pt,opacity=0.3] (e\x) {};}
\draw (e240) node[nodeDecorate,fill=red!50,inner sep=2pt] {};
\draw[lineDecorate,opacity=0.3] (v120) -- (v240);
\draw[lineDecorate,opacity=0.3] (v240) -- (v0);
\draw[lineDecorate,red] (v120) -- (v0);
\draw[lineDecorate,opacity=0.3] (v0) -- (e0);
\draw[lineDecorate,red] (v240) -- (e240);
\draw[lineDecorate,opacity=0.3] (v120) -- (e120);
\end{tikzpicture}}}_{w_{12}-\hat h_3},
\underbrace{
\parbox[c]{30pt}{\begin{tikzpicture}[auto,
  nodeDecorate/.style={shape=circle,inner sep=1pt,draw,thick},%
  lineDecorate/.style={-,thick}]
\draw (0:0.25) node[nodeDecorate] (v0) {\tiny $1$};
\draw (120:0.25) node[nodeDecorate] (v120) {\tiny $2$};
\draw (240:0.25) node[nodeDecorate] (v240) {\tiny $3$};
\foreach \x in {0,120,...,359}{
\draw (\x:0.55) node[nodeDecorate,fill=gray,inner sep=2pt,opacity=0.3] (e\x) {};}
\draw (e0) node[nodeDecorate,fill=red!50,inner sep=2pt] {};
\draw[lineDecorate,red] (v120) -- (v240);
\draw[lineDecorate,opacity=0.3] (v120) -- (v0);
\draw[lineDecorate,opacity=0.3] (v240) -- (v0);
\draw[lineDecorate,opacity=0.3] (v120) -- (e120);
\draw[lineDecorate,opacity=0.3] (v240) -- (e240);
\draw[lineDecorate,red] (v0) -- (e0);
\end{tikzpicture}}}_{w_{23}-\hat h_1}
\right\}\right]\]
For each instance, we have three incoming cavity fields and three weights associated to the cycle edges. In this case there are four possible configurations: one corresponding to the occupation of the external lines only, involving three incoming cavity fields $\hat h$ (larger blobs), and three different configurations in which an edge belonging to the loop is occupied, plus an external edge contributing with its cavity field. The cost is obtained averaging over the minimum-cost configurations.
\subsection{Numerical results}
Let us start discussing the numerical results about the AOC of the MP and the general properties of its scaling in the random-link matching problem on random regular graphs. The results of the cavity computation for different values of $z$ are given in Fig.~\ref{fig:cavitaleading} and in Table~\ref{tab:SummaryTable0}. As expected, with the adopted convention $\lim_{z\to+\infty}E_z=E_\infty$. The cavity estimation has been compared with the AOC obtained solving the problem on random regular graphs with $3\leq z\leq 15$ extracted from $\GRRG(N,z)$ and extrapolating for $N\to+\infty$. For all the investigated values of $z$, the scaling of the AOC is in agreement with the ansatz
\begin{equation}\label{fit}
    E_z(N)=E_z+\frac{E^{(1)}_z}{N}+\frac{E^{(\sfrac{3}{2})}_z}{N^{\sfrac{3}{2}}}+\frac{E^{(2)}_z}{N^2}+o\left(\frac{1}{N^2}\right),
\end{equation}
showing the same anomalous correction appearing in the random-link matching problem on complete graphs \cite{Lucibello2018}, see Fig.~\ref{fig:cavitaleading2}. The numerical estimations obtained extrapolating the asymptotic AOC from finite-$N$ results are in perfect agreement with the cavity prediction for $E_z$. Moreover, we observe that for large values of $z$, the coefficient of the $\sfrac{1}{z}$ term in the large $z$ expansion of $E_z$ is compatible with zero, as expected from the analytical calculation. Assuming a quadratic dependence on $\sfrac{1}{z}$, we found
\begin{equation}
    E_z=\frac{\zeta(2)}{2}+\frac{0.057(1)}{z^2}+o\left(\frac{1}{z^2}\right).
\end{equation}
\begin{table}\centering
\begin{tabular}{@{}ccrr@{}}\toprule
&&$z=3$& $z=4$\\
\midrule
\multirow{2}{*}{Cavity}& $E_z=E^{\mathrm{A}}_z=E_z^{\mathrm{F}}$ & $0.8379047(6)$ & $0.828967(2)$\\
&$E_z^{\mathrm{L}}$ &  $0.6096816(3)$ & $0.65342(1)$ \\
\midrule
\multirow{3}{*}{MP}&
$E_z$ & $0.837903(3)$ & $0.82896(1)$\\
&$E_z^{(1)}$ &  $1.045(3)$ & $0.681(3)$\\
&$E_z^{(\sfrac{3}{2})}$ &  $-3.88(7)$& $-2.72(4)$\\
\midrule
\multirow{3}{*}{AP}&
$E_z^{\mathrm{A}}$ &  $0.83786(7)$ &  $0.828962(7)$\\
&$E_z^{\mathrm{A},(1)}$ &  $-1.5771(3)$ &  $-1.313(2)$\\
&$E_z^{\mathrm{A},(2)}$ &  $7.12(6)$ & $4.2(1)$\\
\midrule
\multirow{3}{*}{FMP}&
$E_z^{\mathrm{F}}$ &  $0.837903(5)$ & $0.828970(1)$\\
&$E_z^{\mathrm{F},(1)}$ &  $-0.03169(6)$ & $-0.0144(3)$\\
&$E_z^{\mathrm{F},(2)}$ &  $-5.60(1)$ & $-3.54(2)$\\
\midrule
\multirow{2}{*}{LMP}&
$E_z^{\mathrm{L}}$ &  $0.609681(6)$ &  $0.653423(1)$\\
&$E_z^{\mathrm{L},(1)}$ &  $0.0154(2)$ & $0.0160(1)$ \\
\bottomrule
\end{tabular}
\caption{Summary table of the numerical estimates for the AOC and its finite-size corrections for the MP and its variants discussed in the paper for $z=3$ and $z=4$. For the details about the numerical simulations and the population dynamics algorithm, see the label of Fig.~\ref{fig:AOC}.
\label{tab:SummaryTable0}}
\end{table}

To analyze now the finite-size correction $E_z^{(1)}$, let us start focusing on the single-cycle contributions. The quantity $\phi_\ell(z)$ has been first evaluated by means of a population dynamics algorithm for different values of $\ell$ and $z$, using Eq.~\eqref{phil1}, and then compared with the results obtained numerically solving the MP on random regular graphs. In particular, the contribution of a cycle of length $\ell$ to the optimal cost can be evaluated numerically using a Markov chain of length $T$ of regular graphs of size $N$, $\{G_t\}_{t=1}^T$, such that $G_{t+1}$ is obtained from $G_t$ by means of a single edge swapping. Let $E_{t}$ be the optimal cost corresponding to the element $G_t$ of the chain, and let $n_\ell^t$ be the number of cycles of length $\ell$ in the graph $G_t$. The average cost shift due to the appearance of a new cycle of length $\ell$ in a graph having $n_\ell$ cycles of length $\ell$ is
\begin{equation}\phi_{\ell}(z;N,T,n)\coloneqq\mediaE{\frac{\sum_{t=1}^{T}\left(E_{t+1}-E_{t}\right)\mathbb I\left(n_\ell^{t+1}=n_\ell^t+1\right)\mathbb I\left(n_\ell^{t}=n\right)}{\sum_{t=1}^{T}\mathbb I\left(n_\ell^{t+1}=n_\ell^t+1\right)\mathbb I\left(n_\ell^{t}=n\right)}}
\end{equation}
The contribution $\phi_\ell(z)$ is obtained numerically extrapolating the previous quantity in the limit of large size and long Markov chains:
\begin{equation}\phi_\ell(z)=
    \lim_{\substack{T\to+\infty\\N\to+\infty}}  \phi_{\ell}(z;N,T,n).\label{phildiretto}
\end{equation}
Note that the right-hand side of the previous equation does not depend on $n$, since the density of cycles of a given length is vanishing in the large size limit. 

\begin{figure}
\centering
    \subfloat[Average value of $\bar n_\ell(z)\phi_\ell(z)$ for $\ell=3,5$ and different values of $z$ (black circles), compared with the numerical estimations (red squares, slightly shifted in the $x$ direction for the sake of clarity). In the inset, zoom of the neighborhood of the origin. \label{costoloop}]{\includegraphics[height=0.44\textwidth]{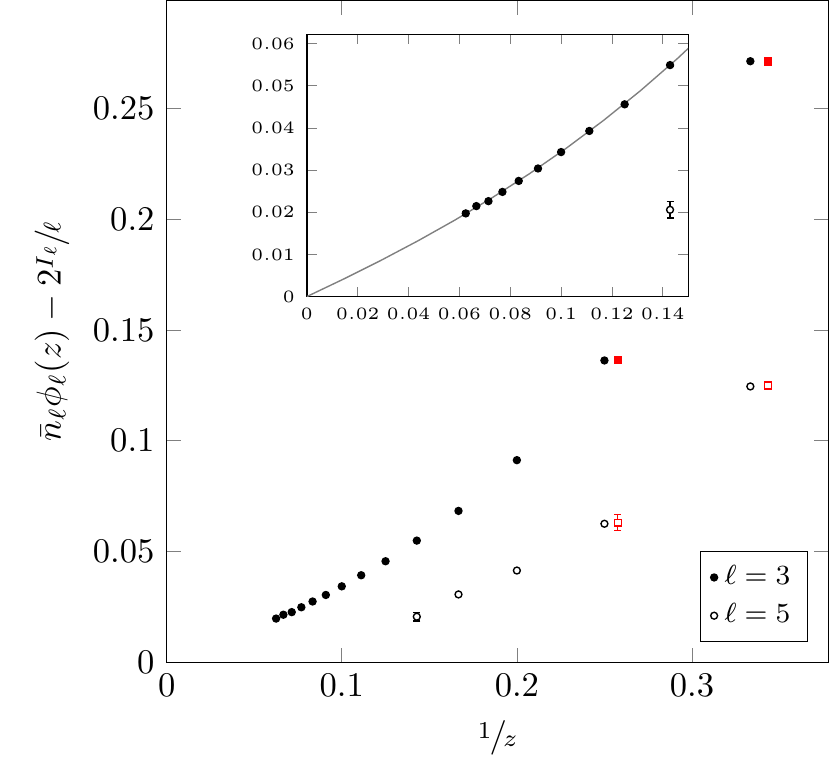}}\hfill
    \subfloat[Value of $\bar n_\ell(3)\phi_\ell(3)$ obtained using the cavity method (black circles) and numerically solving the problem on the graph (red squares, slightly shifted in the $x$ direction for the sake of clarity). See also Fig.~\ref{fig:cavitaloopFr}. \label{fig:cavitaloop}]{\includegraphics[height=0.44\textwidth]{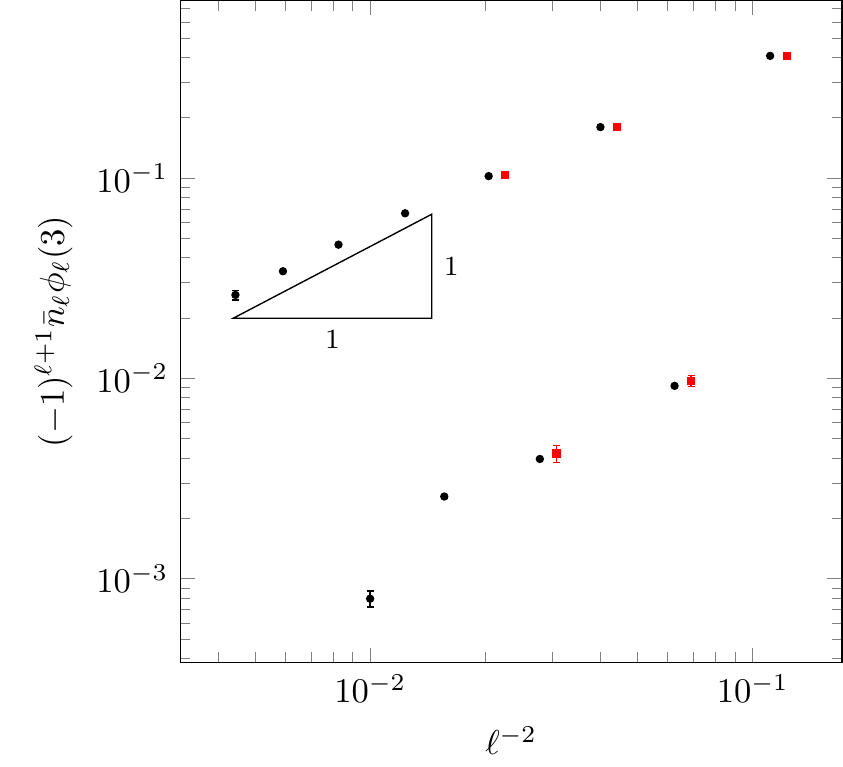}}
 \caption{
 Cavity results for the cycle contributions in the random-link matching problem on random regular graphs. Numerical results obtained solving actual instances of the MP on random regular graphs have been plotted when available. The cavity results have been obtained using a population of $10^8$ fields. Cycle contributions for $\ell>3$ and $z\geq 4$ are very noisy due to the fact that we expect $\phi_\ell(z)$ to scale at least as $(z-1)^{-\ell}$ as $z\to+\infty$ and $\ell\to\infty$. \label{fig:loop}
 }
\end{figure}

By both cavity and direct estimations, we obtained that $|\phi_\ell(z)|$ is decreasing both in $z$ and in $\ell$. The numerical estimations obtained for $z=3$ and $z=4$ and $\ell\leq 5$ are found to be in agreement with the cavity predictions for small values of $\ell$, see Fig.~\ref{fig:cavitaloop} and Table~\ref{tab:SummaryTable}. For larger values of $\ell$, $\ell^{-1}\phi_\ell$ becomes too small and we have not been able to evaluate it numerically solving the MP with statistical significance. On the other hand, we can carry on our cavity estimation to larger cycle lengths. Remarkably, odd cycles and even cycles behaves very differently, and the corresponding contributions lie on different curves, see Fig.~\ref{fig:cavitaloop}. 

\paragraph{Odd-cycles contribution} If $\ell$ is odd, $\phi_\ell(z)$ is positive and, within the precision of our numerical results, we found that $\bar n_\ell(z)\phi_\ell(z)\to 2\ell^{-1}I_\ell$ as $z\to+\infty$ for $\ell=3,5$, see Fig.~\ref{costoloop}. Here $I_\ell$ is the quantity appearing in Eq.~\eqref{costom}: this result supports the hypothesis that the cycles contributions in the MP on random regular graphs give indeed the sum in the finite-size corrections for the MP on the complete graph given in Eq.~\eqref{costom} for large $z$ \cite{Ratieville2002}, a fact a priori not obvious at all due to the fully-connected nature of the latter problem. Moreover, at fixed $z$ and $\ell$ odd, we find that $\bar n_\ell(z)\phi_\ell(z)$ scales as $\ell^{-2}$ for large values of $\ell$. This scaling justifies the presence of the anomalous $N^{-\sfrac{3}{2}}$ correction \cite{Lucibello2017}. Indeed, let us assume that we have a path on the graph of length $k$ arriving to a certain site: the probability that $r$ of the $z-1$ outgoing link go to one of the nodes already occupied by the path is $\binom{z-1}{r}\left(\sfrac{k}{N}\right)^r\left(1-\sfrac{k}{N}\right)^{z-1-r}$. The probability of choosing one of these neighbours is $\frac{r}{z-1}$: from this, it can be seen that the total probability of self-intersecting at the step $k$ equal to $\sfrac{k}{N}$. For an object of final length $\ell$ therefore we have a total probability of self-intersection that scales as $\sfrac{\ell^2}{N}$, that is of order one for $\ell\sim\sqrt{N}$. 
This suggests that we should impose a cut-off to the sum in Eq.~\eqref{sommacicligenerale} of order $\sqrt{N}$. 
But, if the odd-cycle contribution $\bar n_\ell(z)\phi_\ell(z)$ scales as $\ell^{-2}$ for large $\ell$, due to the presence of the cut-off, the odd-cycles part of the sum has a correction that scales as $N^{-\sfrac{1}{2}}$, generating an overall finite-size correction in the AOC that scales as $N^{-\sfrac{3}{2}}$.

\paragraph{Even-cycles contribution} The contribution of even cycles is found to be small (but in general different from zero) and negative, see Fig.~\ref{fig:cavitaloop} and Fig.~\ref{fig:cavitaloopFr} for the values obtained for $z=3$. For $z=4$, using cavity we found $\phi_4(4)=-3.56(4)\cdot 10^{-4}$, to be compared with the numerical estimation $\phi_4(4)=-3.6(4)\cdot 10^{-4}$, whereas the value $\phi_\ell(4)$ for $\ell\geq 6$ was out of the reach of the precision of our numerics. Similarly, the value of $\ell^{-1}\phi_\ell(z)$ for $z>4$ has been found to be smaller than $10^{-5}$ in magnitude, and no significant estimation for it has been obtained, neither using cavity nor the extrapolation in Eq.~\eqref{phildiretto}. We recall here that, in the fully connected case, the sum in Eq.~\eqref{costom} runs over odd contributions only. If the guessed correspondence is true, we expect therefore $\phi_\ell(z)\to 0$ as $z\to \infty$, being this contribution small but finite (and negative) at finite $z$. We will further comment on the even cycle contribution discussing the random fractional matching problem below. 
\begin{table}\centering
\begin{tabular}{@{}crrrr@{}}\toprule
\multirow{2}{*}{$\bar n_{\ell}\phi_\ell(z)$}&\multicolumn{2}{c}{MP}&\multicolumn{1}{c}{FMP}&\multicolumn{1}{c}{LMP}\\
\cmidrule(lr){2-3} \cmidrule(lr){4-4}\cmidrule(lr){5-5}
&$z=3$& $z=4$&$z=3$&$z=3$\\
\midrule
$\ell=3$&$0.4067(1)$&$0.2712(1)$&$-0.005992(4)$&$0.029108(4)$\\
$\ell=4$&$-0.00918(2)$&$-0.00360(4)$&$-0.00918(2)$&$-0.021584(8)$\\
$\ell=5$&$0.1794(3)$&$0.1174(1)$&$-0.00354(1)$&$0.01384(1)$\\
$\ell=6$&$-0.003962(3)$&$-0.0009(3)$&$-0.003962(3)$&$-0.00970(2)$\\
$\ell=7$&$0.10214(5)$&$0.065(9)$&$-0.00192(6)$&$0.00646(3)$\\
$\ell=8$&$-0.00257(4)$&$-0.03(1)$&$-0.00257(4)$&$-0.00449(5)$\\
$\ell=9$&$0.06664(5)$&$0.04(5)$&$-0.0013(3)$&$0.00297(5)$\\
$\ell=10$&$-0.00082(5)$&$0.1(1)$&$-0.0008(6)$&---\\
$\ell=11$&$0.0465(2)$&---&---&---\\
$\ell=13$&$0.0344(4)$&---&---&---\\
$\ell=15$&$0.026(1)$&---&---&---\\
\bottomrule
\end{tabular}
\caption{Summary table of the cavity estimates for $\bar n_\ell(z)\phi_\ell(z)$ for $z=3$ and $z=4$. The results have been obtained using a population of $10^8$ fields.
\label{tab:SummaryTable}}
\end{table}

\paragraph{Wrapping-up} We have therefore verified (term by term) that cycle contributions can be actually computed using cavity, and moreover we have given evidences that these cycle contribution asymptotically converge to the terms in Eq.~\eqref{costom}. We have now to sum up all these contributions to get an estimation of $E_z^{(1)}$. Let us first consider the odd-cyles contribution for $z=3$. We have good cavity results for cycles up to $\ell=15$,
\begin{equation}
\sum_{k=1}^7\bar n_{2k+1}(3)\phi_{2k+1}(3)=0.864(2).
\end{equation}
To estimate the contribution of the remaining odd cycles, we extract the asymptotic behavior of the remaining loops using a quartic polynomial in $\sfrac{1}{\ell}$, obtaining
\begin{equation}
\sum_{k=8}^\infty \bar n_{2k+1}(3)\phi_{2k+1}(3)\simeq 0.2017(8).
\end{equation}
The total contribution of odd cycles is therefore
\begin{equation}\label{disparistandard}
\sum_{k=1}^\infty \bar n_{2k+1}(3)\phi_{2k+1}(3)\simeq 1.065(2).
\end{equation}
A similar estimation can be carried out, in principle, for the even cycle contribution. Unfortunately, the numerics is very noisy, being $\phi_{2k}(3)$ very small for $k>4$. Restricting ourselves to the contributions that we have, we obtain 
\begin{equation}\label{paristandard}
\sum_{k=1}^4 \bar n_{2k+2}(3)\phi_{2k+2}(3)\simeq -0.0165(1).
\end{equation}
Summing the two obtained estimations we have
\begin{equation}
\sum_{k=1}^\infty \bar n_{2k+1}(3)\phi_{2k+1}(3)+\sum_{k=1}^4 \bar n_{2k+2}(3)\phi_{2k+2}(3)\simeq 1.048(2),
\end{equation}
that is compatible with the direct evaluation $E_3^{(1)}=1.045(3)$ obtained via a fit, see Table~\ref{tab:SummaryTable0}. The excellent agreement suggests that the contribution of higher even cycles, and a possible additive constant $\upsilon_3$ not depending on the cycles, is neglectable at our level of precision. The consistency of Eq.~\eqref{sommacicligenerale} can be cross-checked comparing our results with the one corresponding to a variation of the MP that we will discuss below.

\subsection{A note on the assignment problem}
Before studying the fractional matching problem, let us make a comment on the so-called assignment problem. By restricting $\GRRG(N,z)$ to all and only its instances without odd-loops, the MP takes the name of random assignment problem (AP) on a Bethe lattice. The AP has the same leading-order behavior of the MP but, due to the bipartite nature of the graph, the contribution of odd-cycles is absent in the finite-size corrections by construction, whereas the even-cycles contributions coincide with the one evaluated for the MP. However, the ``non-topological'' additive constant $\upsilon_z$ that appears in the finite-size corrections is expected to be different from zero. Indeed, the bipartition of the graph induces a gauge invariance in the cavity equation (in particular, an invariance under a shift of the cavity fields) that produces a nontrivial $\sfrac{1}{N}$ finite-size correction \cite{Mezard1987,Ratieville2002}. Solving the AP on the Bethe lattice, we found no evidence of an $N^{-\sfrac{3}{2}}$ anomalous corrections (as expected if we interpret it as correction to the odd-cycles contribution in the MP), see Fig.~\ref{fig:scalingfract}: the scaling of the AOC with $N$ is therefore exactly of the same type appearing in the fully-connected case \cite{Parisi1998,Linusson2004,Nair2005}.
Because of the arguments given above, the absence of the $N^{-\sfrac{3}{2}}$ anomalous scaling contribution suggests that, for $\ell$ even, $\bar n_\ell\phi_\ell\sim\ell^{-\alpha}$, with $\alpha\geq 3$, or faster, a fact that, however, needs a more accurate numerical investigation. Further information about the AOC in the AP can be found in Table~\ref{tab:SummaryTable0}.

\section{The fractional matching problem}\label{sec:fractional}
The results presented in Ref.~\cite{Lucibello2018} showed that the finite-size corrections in the MP can be better understood relaxing the assumption $m_e\in\{0,1\}$ and studying two variants of the problem, namely the random fractional matching problem and the random ``loopy'' fractional matching problem. In particular, it turns out that the contribution of the sum in Eq.~\eqref{costom} disappears switching from the MP to the fractional matching problem \cite{Wastlund_2010}, suggesting that the finite-size corrections are indeed related to the suppression of cycles in the solution of the MP. We follows this hint for the analysis on sparse topologies.

In the so-called random fractional matching problem (FMP) the occupancy variables $m_e$ are allowed to take any value in $[0,1]$. As in the MP, given a graph $G=(\mathcal V;\mathcal E)$, the cost to be minimized is
\begin{subequations}
\begin{equation}
E^{\mathrm{F}}_G[M]\coloneqq\frac{1}{|M|}\sum_{e\in\mathcal E}w_em_e,
\end{equation}
but this time the constraints read
\begin{equation}
m_e\in[0,1]\quad \forall e\in \mathcal{E}, \quad\sum_{e\rightarrow v}m_e=1\quad \forall v\in \mathcal{V}.
\end{equation}
\end{subequations}
In the equations above we have defined $|M|\coloneqq\sum_{e\in\mathcal{E}}m_e=\sfrac{N}{2}$. As in the MP, we will study the FMP on the random regular graph ensemble $\GRRG(N,z)$, with link weights drawn according to the distribution of Eq.~\eqref{distpesi}. The AOC is then defined as
\begin{equation}
E_z^{\mathrm{F}}(N)\coloneqq\mediaE{\min_M E_G^{\mathrm{F}}[M]}.
\end{equation}
By construction, the average optimal cost $E_z^{\mathrm{F}}(N)$ of the FMP on a Bethe lattice of coordination $z$ satisfies the inequality $E^{\mathrm{F}}_z(N)\leq E_z(N)$. 

\begin{figure}
\centering    
\subfloat[Scaling of the finite-size correction for $z=3$ for the AP and the FMP. See also Table ~\ref{tab:SummaryTable0}. On the $y$-axis we have represented for the FMP $N\Delta E_3^{\mathrm{F}}\coloneqq N(E_3^{\mathrm{F}}(N)-E_3^{\mathrm{F}})$, and similarly for the AP, $N\Delta E_3^{\mathrm{A}}\coloneqq N(E_3^{\mathrm{A}}(N)-E_3^{\mathrm{A}})$, where $E^{\mathrm{A}}_3=\lim_{N\to+\infty}E_3^{\mathrm{A}}(N)$ is the asymptotic AOC of the AP. The continuous lines are quadratic fits in $\sfrac{1}{N}$.\label{fig:scalingfract}]{\includegraphics[height=0.4\textwidth]{
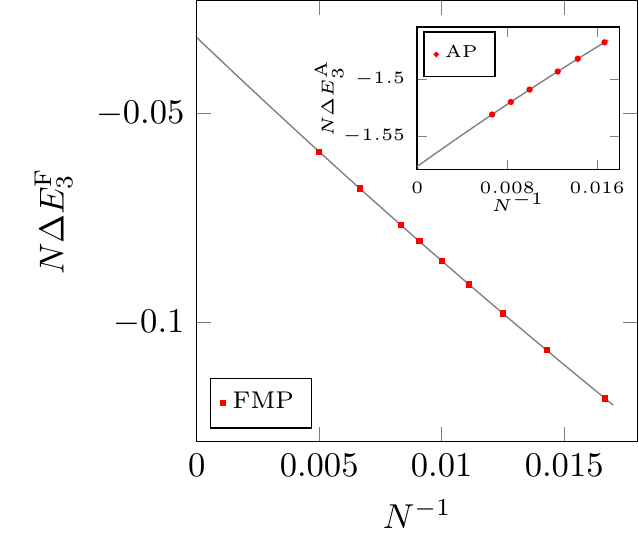}}\hfill
\subfloat[Value of $\bar n_\ell(3)\phi_\ell(3)$ for the FMP obtained using cavity (black circles) and numerically solving the problem (red squares, slightly shifted in the $x$ direction for the sake of clarity). The values for even $\ell$ coincide with the corresponding contribution in the MP. \label{fig:cavitaloopFr}]{\includegraphics[height=0.4\textwidth]{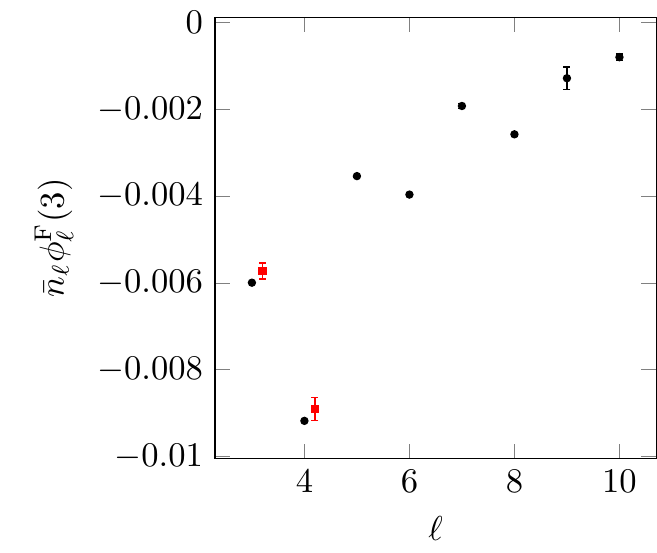}}\hfill

 \caption{
 Cavity and numerical results for the random-link fractional matching problem on random regular graphs with $z=3$.
 }
\end{figure}

It can be proved that, in the optimal solution of a given instance of the FMP, $m_e\in\{0,\sfrac{1}{2},1\}$ \cite{Wastlund_2010}. This result implies that, alongside with isolated edges, cycles and infinite chains might appear in the optimal matching. However, even-length cycles, although possible in principle, cannot be in the optimal solution: given a cycle $\mathcal C$ of even length $\ell$, one of the alternating solutions on it is always cheaper than $\sfrac{1}{2}\sum_{e\in\mathcal C}w_e$.
Using the same argument, it is easily seen that infinite chains of edges with $m_e=\sfrac{1}{2}$ cannot be in the optimal matching as well. It follows that, since on trees the space of feasible solutions of the MP and the FMP is the same, the analysis of Section~\ref{Sec:asymptCost} can be repeated here without any modification, and the leading cost is expected to be identical because of this reason, i.e., $E_z^{\mathrm{F}}\coloneqq \lim_NE_z^{\mathrm{F}}(N)=\lim_NE_z(N)=E_z$. This is indeed the case: this fact also implies that, for large $z$, $E_z^{\mathrm{F}}$ behaves as in Eq.~\eqref{EzMP}, with no $\sfrac{1}{z}$ corrections. A direct study of the scaling of $E_z^{\mathrm{F}}(N)$ with $N$ shows that the MP and the FMP are instead quite different with respect to the scaling of their finite-size corrections: the $N^{-1}$ correction coefficient in the FMP is negative, and there is no $N^{-\sfrac{3}{2}}$ anomalous contribution, see Table~\ref{tab:SummaryTable0} and Fig.~\ref{fig:scalingfract}.

Finite-size corrections are affected by the larger space of feasible solutions in the FMP. We expect the same structure given in Eq.~\eqref{sommacicligenerale} for the MP. Each cycle contribution, however, must be evaluated taking into account that an additional solution $m_e=\sfrac{1}{2}$ for each edge $e$ in the cycle is possible, e.g.,
\[\phi_3^{\mathrm{F},c}(z)=\mathbb E\left[\min\left\{
\parbox[c]{30pt}{\begin{tikzpicture}[auto,
  nodeDecorate/.style={shape=circle,inner sep=1pt,draw,thick},%
  lineDecorate/.style={-,thick}]
\foreach \x in {0,120,...,359}{
\draw (\x:0.25) node[nodeDecorate] (v\x) {};
\draw (\x:0.5) node[nodeDecorate,fill=red!50,inner sep=2pt] (e\x) {};}
\draw[lineDecorate,opacity=0.3] (v0) -- (v120);
\draw[lineDecorate,opacity=0.3] (v240) -- (v0);
\draw[lineDecorate,opacity=0.3] (v240) -- (v120);
\draw[lineDecorate,red] (v120) -- (e120);
\draw[lineDecorate,red] (v240) -- (e240);
\draw[lineDecorate,red] (v0) -- (e0);
\end{tikzpicture}},
\underbrace{\parbox[c]{30pt}{\begin{tikzpicture}[auto,
  nodeDecorate/.style={shape=circle,inner sep=1pt,draw,thick},%
  lineDecorate/.style={-,thick}]
\foreach \x in {0,120,...,359}{
\draw (\x:0.25) node[nodeDecorate] (v\x) {};
\draw (\x:0.5) node[nodeDecorate,fill=gray,inner sep=2pt,opacity=0.3] (e\x) {};}
\draw (e120) node[nodeDecorate,fill=red!50,inner sep=2pt] {};
\draw[lineDecorate,opacity=0.3] (v120) -- (v240);
\draw[lineDecorate,opacity=0.3] (v120) -- (v0);
\draw[lineDecorate,red] (v240) -- (v0);
\draw[lineDecorate,opacity=0.3] (v0) -- (e0);
\draw[lineDecorate,opacity=0.3] (v240) -- (e240);
\draw[lineDecorate,red] (v120) -- (e120);
\end{tikzpicture}}}_{\text{3}},\frac{1}{2}\parbox[c]{30pt}{\begin{tikzpicture}[auto,
  nodeDecorate/.style={shape=circle,inner sep=1pt,draw,thick},%
  lineDecorate/.style={-,thick}]
\foreach \x in {0,120,...,359}{
\draw (\x:0.25) node[nodeDecorate] (v\x) {};
\draw (\x:0.5) node[nodeDecorate,fill=gray,inner sep=2pt,opacity=0.3] (e\x) {};}
\draw[lineDecorate,red] (v240) -- (v120);
\draw[lineDecorate,red] (v240) -- (v0);
\draw[lineDecorate,red] (v0) -- (v120);
\draw[lineDecorate,opacity=0.3] (v0) -- (e0);
\draw[lineDecorate,opacity=0.3] (v120) -- (e120);
\draw[lineDecorate,opacity=0.3] (v240) -- (e240);
\end{tikzpicture}}
\right\}\right]\]
The presence of an additional possible solution does not modify the even-cycles contributions because the ``fractional solution'' will never be the solution of minimum cost, but affects the odd-cycle ones, that, because of the $E_z^{\mathrm{F}}(N)\leq E_z(N)$ inequality, we expect to be smaller than the one appearing in the MP. This has been verified using the cavity method and computing directly the cycle contributions, see Table~\ref{tab:SummaryTable} and Fig.~\ref{fig:cavitaloopFr}: in the FMP all cycle contributions are negative, confirming that switching from the FMP to MP strongly affects the finite-size corrections. The even-cycles contribution of the FMP is identical to the ones appearing in the MP; similarly, the cycle-independent correction $\upsilon_z$, if present, is expected to be the same, because the only modification between the two problems is related to cycles.

We carried on the same analysis performed for the MP for $z=3$, obtaining, for the odd-cycles contribution,
\begin{equation}
\sum_{k=1}^\infty \bar n_{2k+1}(z)\phi_{2k+1}^{\mathrm{F}}(z)\simeq -0.0153(5).
\end{equation}
Adding the even-cycle terms in Eq.~\eqref{paristandard}, we have
\begin{equation}
\sum_{k=1}^\infty \bar n_{2k+1}(3)\phi_{2k+1}^{\mathrm{F}}(3)+\sum_{k=1}^4 \bar n_{2k+2}(3)\phi_{2k+2}(3)\simeq -0.0318(6).
\end{equation}
that is perfectly compatible with the direct numerical estimation $E_z^{\mathrm{F},(1)}=-0.03169(6)$.

Assuming $\upsilon_3$ to be the same in the FMP and in the MP (possibly zero), we can however perform a consistency check that allows us to recover the $\sfrac{1}{N}$ correction of the MP from the odd-cycles contributions of the MP and the total $\sfrac{1}{N}$ correction of the FMP. Observe indeed that the quantity
\begin{equation}
    E_3^{(1)}-E_3^{\mathrm{F},(1)}=1.077(3),
\end{equation}
evaluated using the fit results, should depend on the odd-cycles contributions only, i.e., it should be equal to
\begin{equation}
    \sum_{k=1}^\infty \bar n_{2k+1}(3)\phi_{2k+1}(3)-\sum_{k=1}^\infty \bar n_{2k+1}(3)\phi_{2k+1}^{\mathrm{F}}(3)=1.080(2),
\end{equation}
a number obtained using the cavity results. The two values are in perfect agreement.

\section{The ``loopy'' fractional matching problem}\label{sec:loopy}
The analysis of the finite-size corrections for the fully-connected MP in Ref.~\cite{Lucibello2018} showed that $\sfrac{1}{N}$ corrections disappear entirely in the fully-connected MP with exponential weight distribution if we allow cycles (as in the FMP) and ``self-matched vertices'', i.e. if every vertex can be removed from the graph by paying a price. One might be tempted to perform therefore the same study in the sparse case to get insights on the MP and the FMP, but such a variation can be thought as a modification of the topology of the graph --- due to the addition of $N$ weighted self-loops --- that affects not only the finite-size corrections but also the leading order in $N$. The problem is however interesting by itself and, moreover, we will show that some insight on the fully-connected problem can be obtained anyway. 

To be more precise, in this variant of the FMP, sometimes called random ``loopy'' fractional matching (LMP) \cite{Lucibello2018}, an additional non-negative weight $w_v$ is associated with each vertex $v\in \mathcal{V}$ of the graph. Each $w_v$ is a random variable drawn independently from all other weights with the same probability distribution $\varrho(w)$. An additional ``occupancy variable'' $m_v$ is associated with each vertex. Given a weighted graph $G=(\mathcal V;\mathcal E)$, the target is to search for the set $M=\{m_e\}_{e\in\mathcal E}\cup\{m_v\}_{v\in\mathcal V}$ satisfying the constraints
\begin{subequations}
\begin{equation}
m_e\in[0,1]\quad \forall e\in \mathcal{E},\quad m_v\in[0,1]\quad  \forall v\in \mathcal{V}, \quad\sum_{e\rightarrow v}m_e+ m_v=1\quad \forall v\in \mathcal{V}.
\end{equation}
that maximizes
\begin{equation}
|M|=\sum_e m_e+\frac{1}{2}\sum_v m_v
\end{equation}
and minimizes
\begin{equation}
E^{\mathrm{L}}_G[M]\coloneqq\frac{1}{|M|}\left(\sum_{e\in\mathcal E}w_em_e+  \sum_{v\in\mathcal V}w_vm_v\right).
\end{equation}
We will write for the AOC
\begin{equation}
 E_G^{\mathrm{L}}(N)\coloneqq\mediaE{\min_M E_G^{\mathrm{L}}[M]}.  
\end{equation}
\end{subequations}
Obviously $E_G^{\mathrm{L}}(N)\leq E^{\mathrm{F}}_G(N)\leq E_G(N)$. In other words, in this problem each vertex can be (partially or totally) matched to itself and removed from the graph, paying a cost $m_ew_v$. Note that in this case a matching with $|M|=\sfrac{N}{2}$ always exists. Once again, we are interested in the AOC of the problem in the ensemble of weighted random regular graphs of coordination $z$ and link weights drawn according to the distribution in Eq.~\eqref{distpesi}. 

\subsection{The asymptotic cost}
Similarly to the FMP case, it can be proved that in the optimal configuration $m_e\in\{0,\sfrac{1}{2},1\}$ $\forall e$, and $m_v\in\{0,1\}$ $\forall v$ \cite{Wastlund_2010}, a fact that strongly simplifies the analysis. In addition to this, we can neglect the possibility $m_e=\sfrac{1}{2}$ at the leading order, because of the same arguments given for the FMP for infinite half-occupied chains. Unlike all previously studied problems, the presence of the additional on-vertex degrees of freedom modifies the equation and the AOC at the leading level, i.e., on the tree. In particular, we have two types of marginals on the factor graph, the first $\mu$ going from a vertex to a regular edge, and the second one $\hat\mu$ going from a vertex $i$ to its self-loop $e_i$, occupied by a new variable node
\[\begin{gathered}
\begin{tikzpicture}
\node[fill=black,shape=circle,draw,inner sep=1.5pt] (0) at (0,0) {};
\node[label=above:{},fill=black,shape=circle,position=90:{0.35*\nodeDist} from 0,draw,inner sep=1.5pt] (a2) {};
\node[draw=none,position=30:{0.35*\nodeDist} from a2,inner sep=2pt] (a2a) {};
\node[draw=none,position=150:{0.35*\nodeDist} from a2,inner sep=2pt] (a2b) {};
\node[label=above:{},fill=black,shape=circle,position=270:{0.35*\nodeDist} from 0,draw,inner sep=1.5pt] (a3) {};
\node[draw=none,position=-30:{0.35*\nodeDist} from a3,inner sep=2pt] (a3a) {};
\node[draw=none,position=-150:{0.35*\nodeDist} from a3,inner sep=2pt] (a3b) {};
\draw[thin,gray] (a2a) -- (a2) -- (a2b);
\draw[thin,gray] (a3a) -- (a3) -- (a3b);
\draw[thin,gray] (a2) to [out=60,in=120,looseness=10] (a2);
\draw[thin,gray] (a3) to [out=-60,in=-120,looseness=10] (a3);
\draw[thin] (a3) -- (0) -- (a2);
\node[fill=black,shape=circle, position=0:10mm from 0,draw,inner sep=1.5pt] (i) {};
\node[label=above:{},fill=black,shape=circle,position=90:{0.35*\nodeDist} from i,draw,inner sep=1.5pt] (b4) {};
\node[label=above:{},fill=black,shape=circle,position=-90:{0.35*\nodeDist} from i,draw,inner sep=1.5pt] (b5) {};
\draw[thin] (b4) -- (i) -- (b5);
\node[draw=none,position=30:{0.35*\nodeDist} from b4,inner sep=2pt] (b4a) {};
\node[draw=none,position=150:{0.35*\nodeDist} from b4,inner sep=2pt] (b4b) {};
\node[draw=none,position=-30:{0.35*\nodeDist} from b5,inner sep=2pt] (b5a) {};
\node[draw=none,position=-150:{0.35*\nodeDist} from b5,inner sep=2pt] (b5b) {};
\draw[thin,gray] (b4a) -- (b4) -- (b4b);
\draw[thin,gray] (b5a) -- (b5) -- (b5b);
\draw[thin,gray] (b4) to [out=60,in=120,looseness=10] (b4);
\draw[thin,gray] (b5) to [out=-60,in=-120,looseness=10] (b5);
\draw[thin] (0) -- (i);
\draw (0) to [out=120,in=240,looseness=10] (0);
\draw (i) to [out=60,in=-60,looseness=10] (i);
\end{tikzpicture}
\end{gathered}\Longrightarrow
\begin{gathered}
\begin{tikzpicture}
\node[fill=black,shape=rectangle,draw,inner sep=3pt] (0) at (0,0) {};
\node[shape=circle,draw,inner sep=2pt,label=below:{\small $e$}] (e) at (15mm,0) {};
\node[label=above:{}, fill=gray!60, shape=rectangle, position=90:{0.2\nodeDist} from e, draw,inner sep=2pt] (A) {};
\node[label=above:{},fill=black,shape=rectangle,position=90:{0.7*\nodeDist} from 0,draw,inner sep=2pt] (a2) {};
\node[label=above:{},shape=circle, position=90:{0.2\nodeDist} from 0,draw,inner sep=2pt] (2) {};
\node[label=above:{},fill=black,shape=rectangle,position=270:{0.7*\nodeDist} from 0,draw,inner sep=2pt] (a3) {};
\node[label=above:{}, shape=circle, position=103:{0.75\nodeDist} from a3] (label) {\small $i$};
\node[label=above:{},shape=circle, position=270:{0.2\nodeDist} from 0,draw,inner sep=2pt] (3) {};
\node[fill=red!40,shape=circle, position=180:{2.365\nodeDist} from 0,draw,inner sep=2pt,label=below:{\small $e_i$}] (v1) {};
\node[label=above:{}, fill=gray!60, shape=rectangle, position=180:{0.2\nodeDist} from v1, draw,inner sep=2pt] (v1f) {};
\draw[thin] (v1f) -- (v1);
\path (v1) edge [<-,>=stealth'] node [fill=white] {\small $\hat\mu^{i\to e_i}$} (0) 
(0) edge [->,>=stealth'] node [fill=white] {\small $\mu^{i\to e}$}  (e);
\node[label=above:{}, fill=gray!60, shape=rectangle, position=0:{0.2\nodeDist} from 2, draw,inner sep=2pt] (A2) {};
\node[label=above:{}, fill=gray!60, shape=rectangle, position=0:{0.2\nodeDist} from 3, draw,inner sep=2pt] (A3) {};
\draw[thin] (0) -- (2) -- (a2) -- (2) -- (A2);
\draw[thin] (0) -- (3) -- (a3) -- (3) -- (A3);
\node[fill=black, position=0:5mm from e,draw,inner sep=3pt] (i) {};
\node[label=above:{},fill=black,shape=rectangle,position=90:{0.7*\nodeDist} from i,draw,inner sep=2pt] (b4) {};
\node[label=above:{},shape=circle, position=90:{0.2\nodeDist} from i,draw,inner sep=2pt] (4) {};
\node[label=above:{}, fill=gray!60, shape=rectangle, position=180:{0.2\nodeDist} from 4, draw,inner sep=2pt] (B4) {};
\node[label=above:{},fill=black,shape=rectangle,position=-90:{0.7*\nodeDist} from i,draw,inner sep=2pt] (b5) {};
\node[label=above:{},shape=circle, position=-90:{0.2*\nodeDist} from i,draw,inner sep=2pt] (5) {};\node[label=above:{}, fill=gray!60, shape=rectangle, position=-180:{0.25\nodeDist} from 5, draw,inner sep=2pt] (B5) {};
\node[fill=red!40,shape=circle, position=0:{0.2\nodeDist} from i,draw,inner sep=2pt] (v2) {};
\node[label=above:{}, fill=gray!60, shape=rectangle, position=0:{0.2\nodeDist} from v2, draw,inner sep=2pt] (v2f) {};
\draw[thin] (i) -- (4) -- (b4) -- (4) -- (B4);
\draw[thin] (v2f) -- (v2) -- (i) -- (e) -- (A);
\draw[thin] (i) -- (5) -- (b5) -- (5) -- (B5);
\end{tikzpicture}
\end{gathered}\]
Let us now parametrize $\mu$ and $\hat\mu$ in such a way that, at finite temperature,
\begin{equation}
h^{i\to e}=-\frac{1}{\beta}\ln\frac{\mu^{i\to e}(0)}{\mu^{i\to e}(1)},
\quad s_i=-\frac{1}{\beta}\ln\frac{\hat \mu^{i\to e_i}(0)}{\hat \mu^{i\to e_i}(1)}.
\end{equation}
For $\beta\to+\infty$ we have
\begin{subequations}\label{cavityeqloopy}
\begin{align}
s&\stackrel{\mathrm d}{=}\min_{{1\leq e\leq z}}\left(w_e-h_e\right)
\\
h&\stackrel{\mathrm d}{=}\min\left\{\min_{{1\leq e\leq z-1}}\left(w_e-h_e\right),w_v\right\}.
\end{align}
\end{subequations}
For $w_v\to+\infty$ the cavity equations for the MP/FMP are recovered.  If we denote by $p_z(h)$ the distribution of $h$ and by $\hat p_z(s)$ the distribution of $s$, the AOC is given by
\begin{multline}\label{costoloopycav}
E_z^{\mathrm{L}}=z\int\dd w\,w\varrho_z(w)\iint\dd h\dd h^{\prime}\theta\left(h+h^{\prime}-w\right)\,p_z(h)p_z(h^{\prime})\\
+2\int\dd w\,w\varrho_z(w)\int\dd s\,\hat p_z(s)\,\theta(s-w).
\end{multline}
The asymptotic AOC $E_z^{\mathrm{L}}\neq E_z$ is found to satisfy a different scaling relation in $z$ with respect to the MP and the FMP. In particular, we find
\begin{equation}\label{costoloopyfit}
E_z^{\mathrm{L}}=\frac{\zeta(2)}{2}-\frac{\zeta(2)}{2z}+o\left(\frac{1}{z}\right).
\end{equation}
i.e., a negative linear term in $\sfrac{1}{z}$ is present. Eq.~\eqref{costoloopyfit} is proven in the Appendix. We numerically verified the cavity predictions for $3\leq z\leq 15$, see Fig.~\ref{fig:loopyA}. The $\sfrac{1}{z}$ correction at the leading order is due to the presence of self-loops, that are occupied with finite probability. The second integral in Eq.~\eqref{costoloopycav}, in particular, corresponds to the average self-loop cost $\phi_1^{\mathrm{L}}(z)$. Observing that $s$ and $h$ are identically distributed for $z\to+\infty$, it can be evaluated exactly for $z\gg 1$ as 
\begin{equation}\label{costoselfloops}
    \phi_1^{\mathrm{L}}(z)\coloneqq \int\dd w\,w\varrho_z(w)\int\dd s\,\hat p_z(s)\theta(s-w)=\frac{\zeta(2)}{4z}+o\left(\frac{1}{z}\right),
\end{equation}
see Fig.~\ref{fig:loopyE}. The probability that a vertex is ``self-matched'', on the other hand, decreases as (see Fig.~\ref{fig:loopyE})
\begin{equation}
p_s(z)=\int\dd w\, \varrho_z(w)\int\dd s\,\hat p_z(s)\theta(s-w)=\frac{\ln 2}{z}+o\left(\frac{1}{z}\right).
\end{equation}

\subsection{Finite-size corrections}
\begin{figure}
\centering    
\subfloat[AOC for the LMP as function of $z$. The asymptotic correction to the mean field value scales as $z^{-1}$. Numerical simulations (red squares, shifted by $10^{-2}$ in the $x$ direction for the sake of clarity) are in perfect agreement with the cavity prediction (black circles). A cubic fit in $\sfrac{1}{z}$ is also represented. In the inset, cost density of the LMP for $z=3$ as a function of $\sfrac{1}{N}$. The continuous line is a quadratic fit in $\sfrac{1}{N}$.
\label{fig:loopyA}]{\includegraphics[height=0.32\textwidth]{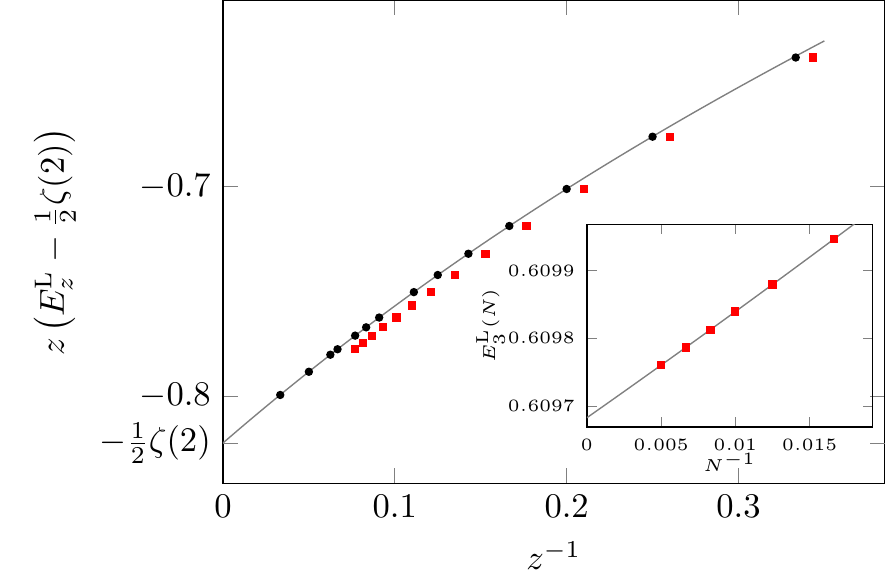}}\hfill
\subfloat[Cavity estimation of the average cost $\phi_1^{\mathrm{L}}(z)$ of a self-loop as a function of $z$ in the LMP. In the inset, occupation probability $p_{s}(z)$ of the self-loop as a function of $z$. Note that the asymptotic value is given by $\lim_{z\to\infty}zp_s(z)=\ln 2$ and higher order corrections scale as $\sfrac{1}{z^2}$.\label{fig:loopyE}]{\includegraphics[height=0.33\textwidth]{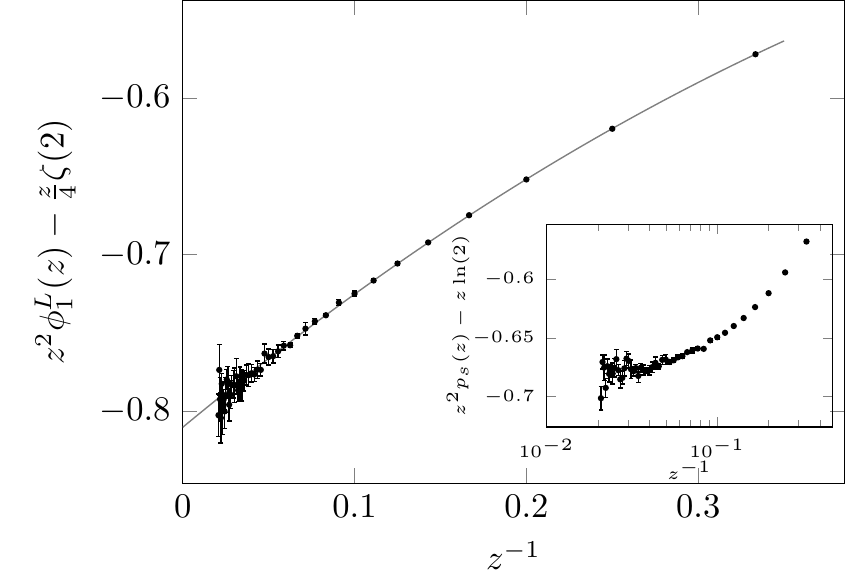}}\\
\subfloat[Value of $\bar n_\ell(3)\phi_\ell(3)$ for the LMP. Numerical results (red squares, shifted in the $x$ direction for the sake of clarity) are compared with the cavity predictions (black circles). The observed decaying is expontial. The gray curves corresponds to an exponential fit of the tipe $f(\ell)=ae^{-b\ell}$, with $a=0.094(4)$ and $b=0.381(5)$. \label{fig:loopyC}]{\includegraphics[height=0.33\textwidth]{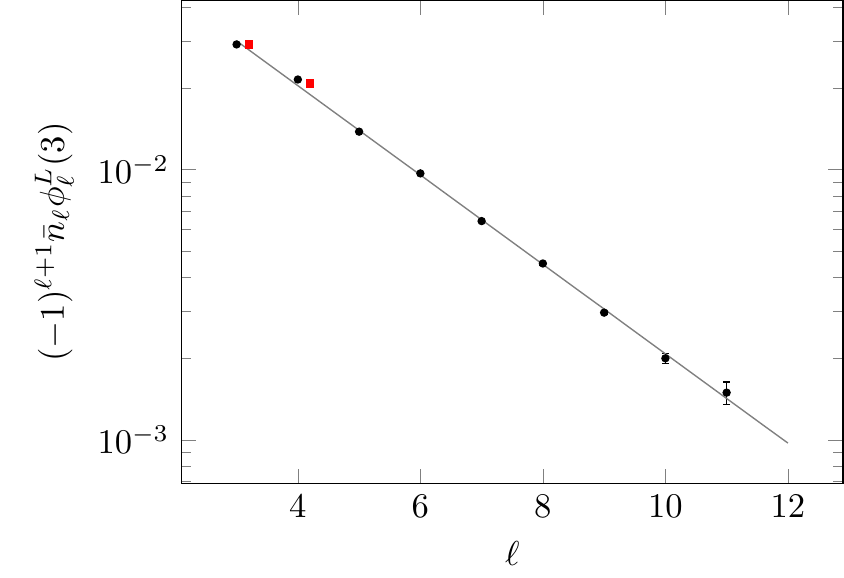}}
\hfill
\subfloat[Loop contribution for $\ell=3$ as function of $z$ in the LMP obtained using the cavity method.\label{fig:loopyD}]{\includegraphics[height=0.33\textwidth]{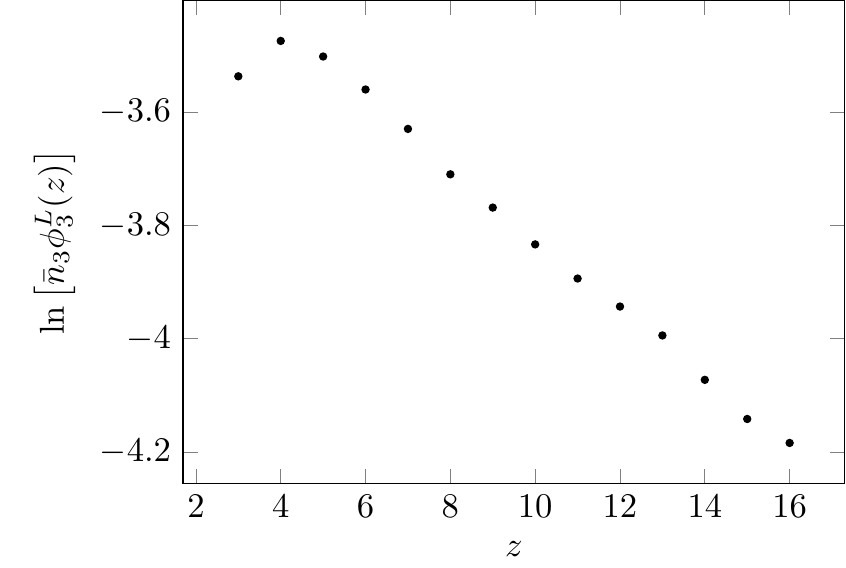}}

\caption{Cavity and numerical results for the random loopy fractional matching problem.}
\end{figure}

Finite-size corrections are also expected to be affected by the larger space of feasible solutions. The LMP has no anomalous corrections, see Fig.~\ref{fig:loopyA}, as in the FMP. However, both even and odd cycle contributions are different with respect to the corresponding ones in the MP and the FMP, see Table~\ref{tab:SummaryTable}. As in the previous cases, each cycle contribution must be evaluated taking into account all allowed configurations in the FMP, plus the possibilities that vertices are self-matched. Pictorially,
\[\phi_3^{\mathrm{L},c}(z)=\mathbb E\left[\min\left\{
\parbox[c]{35pt}{\begin{tikzpicture}[auto,
  nodeDecorate/.style={shape=circle,inner sep=1pt,draw,thick},%
  lineDecorate/.style={-,thick}]
\foreach \x in {0,120,...,359}{
\draw (\x:0.35) node[nodeDecorate] (v\x) {};
\draw (\x:0.6) node[nodeDecorate,fill=red!50,inner sep=2pt] (e\x) {};}
\draw[lineDecorate,opacity=0.3] (v0) -- (v120);
\draw[lineDecorate,opacity=0.3] (v240) -- (v0);
\draw[lineDecorate,opacity=0.3] (v240) -- (v120);
\draw[lineDecorate,red] (v120) -- (e120);
\draw[lineDecorate,red] (v240) -- (e240);
\draw[lineDecorate,red] (v0) -- (e0);
\draw[lineDecorate,opacity=0.3] (v0) to [out=150,in=210,looseness=15] (v0);
\draw[lineDecorate,opacity=0.3] (v120) to [out=270,in=330,looseness=15,opacity=0.3] (v120);
\draw[lineDecorate,opacity=0.3] (v240) to [out=30,in=90,looseness=15] (v240);
\end{tikzpicture}},
\underbrace{\parbox[c]{35pt}{\begin{tikzpicture}[auto,
  nodeDecorate/.style={shape=circle,inner sep=1pt,draw,thick},%
  lineDecorate/.style={-,thick}]
\foreach \x in {0,120,...,359}{
\draw (\x:0.35) node[nodeDecorate] (v\x) {};
\draw (\x:0.6) node[nodeDecorate,fill=gray,inner sep=2pt,opacity=0.3] (e\x) {};}
\draw (e120) node[nodeDecorate,fill=red!50,inner sep=2pt] {};
\draw[lineDecorate,opacity=0.3] (v120) -- (v240);
\draw[lineDecorate,opacity=0.3] (v120) -- (v0);
\draw[lineDecorate,red] (v240) -- (v0);
\draw[lineDecorate,opacity=0.3] (v0) -- (e0);
\draw[lineDecorate,opacity=0.3] (v240) -- (e240);
\draw[lineDecorate,red] (v120) -- (e120);
\draw[lineDecorate,opacity=0.3] (v0) to [out=150,in=210,looseness=15] (v0);
\draw[lineDecorate,opacity=0.3] (v120) to [out=270,in=330,looseness=15,opacity=0.3] (v120);
\draw[lineDecorate,opacity=0.3] (v240) to [out=30,in=90,looseness=15] (v240);
\end{tikzpicture}}}_{\text{3}},
\frac{1}{2}\parbox[c]{35pt}{\begin{tikzpicture}[auto,
  nodeDecorate/.style={shape=circle,inner sep=1pt,draw,thick},%
  lineDecorate/.style={-,thick}]
\foreach \x in {0,120,...,359}{
\draw (\x:0.35) node[nodeDecorate] (v\x) {};
\draw (\x:0.6) node[nodeDecorate,fill=gray,inner sep=2pt,opacity=0.3] (e\x) {};}
\draw[lineDecorate,red] (v240) -- (v120);
\draw[lineDecorate,red] (v240) -- (v0);
\draw[lineDecorate,red] (v0) -- (v120);
\draw[lineDecorate,opacity=0.3] (v0) -- (e0);
\draw[lineDecorate,opacity=0.3] (v240) -- (e240);
\draw[lineDecorate,opacity=0.3] (v120) -- (e120);
\draw[lineDecorate,opacity=0.3] (v0) to [out=150,in=210,looseness=15] (v0);
\draw[lineDecorate,opacity=0.3] (v120) to [out=270,in=330,looseness=15,opacity=0.3] (v120);
\draw[lineDecorate,opacity=0.3] (v240) to [out=30,in=90,looseness=15] (v240);
\end{tikzpicture}},
\parbox[c]{35pt}{\begin{tikzpicture}[auto,
  nodeDecorate/.style={shape=circle,inner sep=1pt,draw,thick},%
  lineDecorate/.style={-,thick}]
\foreach \x in {0,120,...,359}{
\draw (\x:0.35) node[nodeDecorate] (v\x) {};
\draw (\x:0.6) node[nodeDecorate,fill=gray,inner sep=2pt,opacity=0.3] (e\x) {};}
\draw[lineDecorate,opacity=0.3] (v240) -- (v120);
\draw[lineDecorate,opacity=0.3] (v240) -- (v0);
\draw[lineDecorate,opacity=0.3] (v0) -- (v120);
\draw[lineDecorate,opacity=0.3] (v0) -- (e0);
\draw[lineDecorate,opacity=0.3] (v240) -- (e240);
\draw[lineDecorate,opacity=0.3] (v120) -- (e120);
\draw[lineDecorate,red] (v0) to [out=150,in=210,looseness=15] (v0);
\draw[lineDecorate,red] (v120) to [out=270,in=330,looseness=15] (v120);
\draw[lineDecorate,red] (v240) to [out=30,in=90,looseness=15] (v240);
\end{tikzpicture}},\underbrace{\parbox[c]{35pt}{\begin{tikzpicture}[auto,
  nodeDecorate/.style={shape=circle,inner sep=1pt,draw,thick},%
  lineDecorate/.style={-,thick}]
\foreach \x in {0,120,...,359}{
\draw (\x:0.35) node[nodeDecorate] (v\x) {};
\draw (\x:0.6) node[nodeDecorate,fill=gray,inner sep=2pt,opacity=0.3] (e\x) {};}
\draw[lineDecorate,red] (v240) -- (v120);
\draw[lineDecorate,opacity=0.3] (v240) -- (v0);
\draw[lineDecorate,opacity=0.3] (v0) -- (v120);
\draw[lineDecorate,opacity=0.3] (v0) -- (e0);
\draw[lineDecorate,opacity=0.3] (v240) -- (e240);
\draw[lineDecorate,opacity=0.3] (v120) -- (e120);
\draw[lineDecorate,red] (v0) to [out=150,in=210,looseness=15] (v0);
\draw[lineDecorate,opacity=0.3] (v120) to [out=270,in=330,looseness=15] (v120);
\draw[lineDecorate,opacity=0.3] (v240) to [out=30,in=90,looseness=15] (v240);
\end{tikzpicture}}}_{\text{3}},\underbrace{\parbox[c]{35pt}{\begin{tikzpicture}[auto,
  nodeDecorate/.style={shape=circle,inner sep=1pt,draw,thick},%
  lineDecorate/.style={-,thick}]
\foreach \x in {0,120,...,239}{
\draw (\x:0.35) node[nodeDecorate] (v\x) {};
\draw (\x:0.6) node[nodeDecorate,fill=gray,inner sep=2pt,opacity=0.3] (e\x) {};}
\draw (240:0.35) node[nodeDecorate] (v240) {};
\draw (240:0.6) node[nodeDecorate,fill=red!50,inner sep=2pt] (e240) {};
\draw[lineDecorate,opacity=0.3] (v240) -- (v120);
\draw[lineDecorate,opacity=0.3] (v240) -- (v0);
\draw[lineDecorate,opacity=0.3] (v0) -- (v120);
\draw[lineDecorate,opacity=0.3] (v0) -- (e0);
\draw[lineDecorate,red] (v240) -- (e240);
\draw[lineDecorate,opacity=0.3] (v120) -- (e120);
\draw[lineDecorate,red] (v0) to [out=150,in=210,looseness=15] (v0);
\draw[lineDecorate,red] (v120) to [out=270,in=330,looseness=15] (v120);
\draw[lineDecorate,opacity=0.3] (v240) to [out=30,in=90,looseness=15] (v240);
\end{tikzpicture}}}_{\text{3}},
\underbrace{\parbox[c]{35pt}{\begin{tikzpicture}[auto,
  nodeDecorate/.style={shape=circle,inner sep=1pt,draw,thick},%
  lineDecorate/.style={-,thick}]
\draw (120:0.35) node[nodeDecorate] (v120) {};
\draw (120:0.6) node[nodeDecorate,fill=gray,inner sep=2pt,opacity=0.3] (e120) {};
\draw (0:0.35) node[nodeDecorate] (v0) {};
\draw (0:0.6) node[nodeDecorate,fill=red!50,inner sep=2pt] (e0) {};
\draw (240:0.35) node[nodeDecorate] (v240) {};
\draw (240:0.6) node[nodeDecorate,fill=red!50,inner sep=2pt] (e240) {};
\draw[lineDecorate,opacity=0.3] (v240) -- (v120);
\draw[lineDecorate,opacity=0.3] (v240) -- (v0);
\draw[lineDecorate,opacity=0.3] (v0) -- (v120);
\draw[lineDecorate,red] (v0) -- (e0);
\draw[lineDecorate,red] (v240) -- (e240);
\draw[lineDecorate,opacity=0.3] (v120) -- (e120);
\draw[lineDecorate,opacity=0.3] (v0) to [out=150,in=210,looseness=15] (v0);
\draw[lineDecorate,red] (v120) to [out=270,in=330,looseness=15] (v120);
\draw[lineDecorate,opacity=0.3] (v240) to [out=30,in=90,looseness=15] (v240);
\end{tikzpicture}}}_{3}
\right\}\right]\]
As in the FMP, $\bar n_\ell(z)\phi_\ell^{\mathrm{L}}(z)$ decays to zero very fast both in $\ell$ and in $z$, making the numerical evaluation for $z\geq 4$ and $\ell\geq 4$ very difficult. In Fig.~\ref{fig:loopyD} we present our cavity results for $\bar n_3\phi_3^{\mathrm{L}}(z)$, that appears to go to zero exponentially fast in $z$. For $z=3$, the computation is less cumbersome: the cavity results for the corrections in the $z=3$ case are given in Table~\ref{tab:SummaryTable}, and in Fig.~\ref{fig:loopyC}, where they are compared with the numerical estimations obtained for $\ell=3$ and $\ell=4$. As in the MP, $\phi^{\mathrm{L}}_\ell(3)$ is found to be positive for odd cycles and negative for even cycles. The sum of the first contributions is
\begin{equation}
    \sum_{\ell=3}^9\bar n_\ell(3)\phi^{\mathrm{L}}_\ell(3)=0.0166(2),
\end{equation}
that is close, but non compatible, with the result obtained from a fit, given in Table~\ref{tab:SummaryTable0}, $E_z^{\mathrm{L},(1)}=0.0154(2)$, due to the fact that all contributions for $\ell>9$ have been neglected. The quantity $|\bar n_\ell\phi^{\mathrm{L}}_\ell(3)|$ is found to go to zero exponentially fast, see Fig.~\ref{fig:loopyC}. Extrapolating through the fit parameters, we estimate
\begin{equation}
    \sum_{\ell=3}^\infty \bar n_\ell(3)\phi^{\mathrm{L}}_\ell(3)\simeq 0.0153(5),
\end{equation}
that is compatible with the fit result. As in the MP and the FMP, no additional constant is found within the precision of our calculations.

\section{The fully-connected limit}\label{sec:zN}
As anticipated, the obtained results shed some light on the nature of the finite-size corrections in the fully-connected model. With the exception of the odd-cycles contribution in the MP, we have observed that all cycle corrections go to zero as $z\to +\infty$ in all considered models, at least within our numerical precision. Our results suggest that, for large $N$
\begin{equation}
    \lim_{z\to N-1}E_z(N)=\frac{\zeta(2)}{2}+\frac{1}{N}\sum_{k=1}^\infty\lim_{z\to\infty}\bar n_{k}(z)\phi_{k}(z)+o\left(\frac{1}{N}\right)
    =\frac{\zeta(2)}{2}+\frac{2}{N}\sum_{k=1}^\infty\frac{I_{2k+1}}{2k+1}+o\left(\frac{1}{N}\right).
\end{equation}
Similarly, for the FMP, no $\sfrac{1}{N}$ finite-size correction is expected for $z\to N-1$.

These predictions, based on the assumption that the sparse-dense correspondence holds also at the level of finite-size corrections, are equal to the exact results obtained on the complete graph $K_N$ given in Ref.~\cite{Lucibello2018}, \textit{up to an additional correction}, that is the same in all cases and equal to $\sfrac{\zeta(2)}{2N}$, as can be seen from Eq.~\eqref{costom} with reference to the MP. This apparent disagreement is however only due to a different choice of the weight distribution density. To better understand the subtlety behind this apparent discrepancy, let us restrict to the MP. In Ref.~\cite{Lucibello2018}, the authors consider the cost function $\hat E_{K_N}[M]$ that is slightly different from the one in Eq.~\eqref{costoEG}, i.e.,
\begin{equation}
    \hat E_{K_N}[M]=\sum_{e}m_e \hat w_e=\frac{2}{N-1}\sum_{e} m_ew_e=\frac{N}{N-1}E_{K_N}[M],
\end{equation}
where the quantities $\hat w_e$ are i.i.d.~random variable extracted from $\hat \varrho(\hat w)=\e^{-\hat w}$, and we have performed the change of variable $w=\frac{N-1}{2}\hat w$, so that $\hat w$ are extracted from $\varrho_{N-1}(w)$, as should be in our convention. By consequence, the cost in Ref.~\cite{Lucibello2018} is such that
\begin{equation}
    \hat E_{K_N}=\frac{N}{N-1}E_{K_N}=E_{K_N}+\frac{E_{K_N}}{N}+o\left(\frac{1}{N}\right).
\end{equation}
For the same reason, this correction appears both in the FMP and in the LMP. 

Aside for this contribution, that only appears because of different conventions in the definition of the cost, the fully connected case is recovered taking $z\to N-1$. For finite $z$, the leading term $E_z$ in the MP and the FMP has no $\sfrac{1}{z}$ correction, so taking $z\to N-1$ does not give additional $\sfrac{1}{N}$ terms, and the fully-connected limit is recovered. In the LMP, instead, a $\sfrac{1}{z}$ correction is present, see Eq.~\eqref{costoloopyfit}. Taking $z\to N-1$ and observing that the cycles contribution goes to zero for $z=N-1\to+\infty$, we have
\begin{equation}
    \lim_{z\to N-1}E_z^\text{L}(N)=\frac{\zeta(2)}{2}-\frac{\zeta(2)}{2N}+o\left(\frac{1}{N}\right).
\end{equation}
In particular, adopting the convention of Ref.~\cite{Lucibello2018}, the additional contribution exactly cancels the $\sfrac{1}{N}$ correction above, giving no $\sfrac{1}{N}$ terms.

\section{Conclusions}\label{sec:conclusioni}
In the present work we have analyzed the random-link matching problem on random regular graphs of coordination $z$, alongside with two variants of the problem, namely the fractional matching problem and the ``loopy'' fractional matching problem. In all cases, the asymptotic average optimal cost has been computed using the cavity method, and compared with numerical results obtained solving a large number of instances of the corresponding problem. In the spirit of previous finite-size analyses of disordered systems on sparse topologies, we also evaluated the finite-size corrections, assuming that they can be decomposed in contributions of single topological structures (here cycles) appearing in the graph with density $O(\sfrac{1}{N})$. Due to the fact that each cycle of length $\ell$ can be thought as embedded in an infinite tree, its average contribution to the cost $\phi_\ell(z)$ can be evaluated, once again, by means of the cavity method, and must then be re-weighted by its multiplicity $\bar n_\ell(z)$. The quantities $\phi_\ell(z)$ go to zero as $z^{-\ell}$, or faster, for large values of $z$ and $\ell$: the analysis has been carried on therefore mostly for $z=3$. Our results can be summarized as follows.

In the random-link matching problem, odd cycles and even cycles are found to contribute very differently to the finite-size corrections. Odd-cycles terms $\bar n_\ell\phi_\ell$ are positive and scale as $\ell^{-2}$ for large $\ell$, implying the presence of an additional anomalous $N^{-\sfrac{3}{2}}$ correction, that is indeed found in our numerical results. Even-cycles contributions, on the other hand, are found to be negative and smaller in modulus, with a faster decay in $\ell$ and infinitesimal for large $z$ (the very fast decay in $z$ and the small modulus did not allow us to extract the decaying properly). We have also found a strong evidence that, for $\ell=3$ and $\ell=5$, $\bar n_\ell\phi_\ell$ converges to a corresponding ``odd-cycle-like'' contribution appearing in the finite-size corrections of the fully-connected problem. The correspondence of the fully-connected finite-size expansion with actual cycles in the graph had been suggested in Refs.~\cite{Ratieville2002,Lucibello2017,Lucibello2018}, but not directly verified: our result support this claim and the consequent argument given in Ref.~\cite{Lucibello2017} for the anomalous correction appearing on the complete graph. Remarkably, the cavity prediction for each cycle contribution is in agreement with the numerical estimation obtained solving directly the problem, whenever such an estimation has been practicable. The sum of all cycle contributions evaluated using the cavity method is fully compatible with our numerical results for the $\sfrac{1}{N}$ correction in the random-link matching problem. In this way, our ansatz for the finite-size correction has been verified both globally and term by term.

As additional cross-check, we also considered the so called fractional matching problem on random regular graphs. In this version of the matching problem, the optimal configuration can be made of both odd cycles and dimers. We have found that the fractional matching and the regular one are asymptotically equivalent for any value of the coordination $z$, but they differ in the finite-size corrections. As expected, only the odd-cycles corrections are found to be different with respect to the standard matching, their scaling with $\ell$ being faster than $\ell^{-3}$: this implies no $N^{-\sfrac{3}{2}}$ correction in the fractional matching, a fact that has been verified numerically. Once again, within the precision of our cavity results, the finite-size corrections found numerically are compatible with our cycle expansion.

Finally, we have studied, in the same way, the so-called ``loopy'' fractional matching problem, that is a variant of the fractional matching problem in which a vertex can be removed paying a price of ``self-matching''. For finite $z$, vertices are self-matched with finite probability. Both the leading cost and the finite-size correction can be estimated using the cavity method, the latter by means of the usual cycle expansion.

Using the cavity method, we have computed analytically the $O(\sfrac{1}{z})$ corrections to the asymptotic average optimal cost in all discussed problems. Moreover, with the exception of the assignment problem, we have found no numerical evidence of additional cycle-independent contributions, although they cannot be excluded \textit{a priori}. Such contributions might in general appear with a different choice of the weight distribution $\varrho$, as it happens on the complete graph \cite{Ratieville2002,Lucibello2018}.

\section*{Acknowledgments}
The authors would like to thank Enzo Marinari, Federico Ricci-Tersenghi and Tommaso Rizzo for useful discussions. We also thank the anonymous referees for the careful reading of the manuscript and valuable suggestions that improved the readability of the paper. The research of GP and GS has been supported by the Simons Foundation (grant No.~454949, G.~Parisi).
\appendix
\section{Finite-$z$ corrections}
\begin{figure}
\centering    
\subfloat[Plot of the $\sfrac{1}{z}$ correction to the $z\to+\infty$ cavity field probability distribution for the MP and the FMP.\label{fig:DistribuzCorrStandard}]{\includegraphics[width=0.48\textwidth]{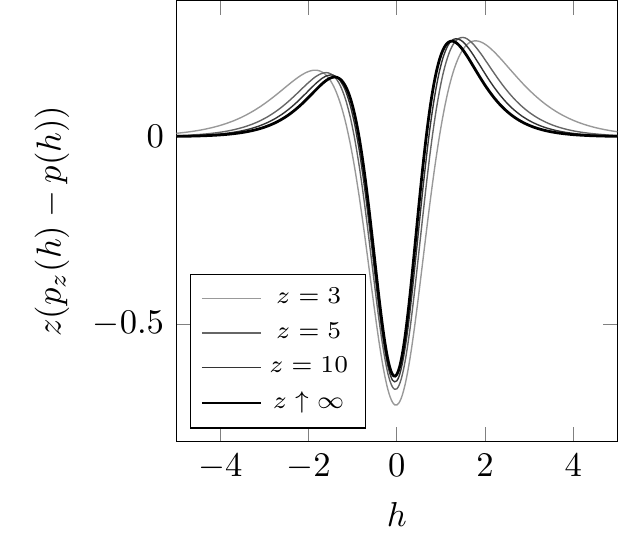}}\hfill
\subfloat[Plot of the $\sfrac{1}{z}$ correction to the $z\to+\infty$ cavity field probability distribution for the LMP. \label{fig:DistribuzCorrLoopy}]{\includegraphics[width=0.5\textwidth]{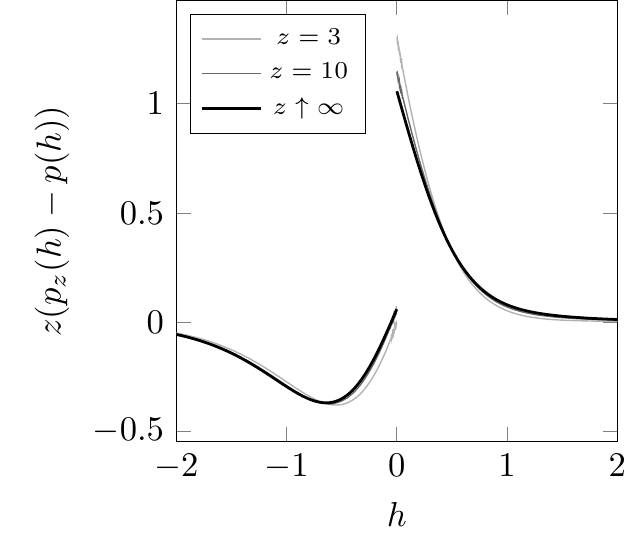}}
\caption{Plots of the $\sfrac{1}{z}$ corrections to the cavity fields distribution in the analyzed random matching problems. The finite-$z$ results have been obtained using a population dynamics algorithm.}
\end{figure}

In this Appendix we study Eq.~\eqref{cavitaPhiz} for large values of $z$. Let us start denoting by
\begin{equation}
    \Phi_z(x)\coloneqq \int_{x}^\infty p_z(h)\dd h=\Phi(x)+\frac{\Phi^{(1)}(x)}{z}+o\left(\frac{1}{z}\right),
\end{equation}
where $\Phi(x)=\lim_{z\to+\infty}\Phi_z(x)$. The equation for $\Phi(x)$ is obtained from Eq.~\eqref{cavitaPhiz} taking $z\to\infty$,
\begin{equation}
\Phi(x)=\exp\left(-2\int_0^\infty \Phi(w-x)\dd w\right),
\end{equation}
that has solution \cite{Mezard1985}
\begin{equation}
\label{distribCavAsintotica}
\Phi(x)=\lim_{z\to+\infty}\Phi_z(x)=\frac{1}{1+\e^{2x}}\Rightarrow p(h)=\frac{2\e^{2h}}{(1+\e^{2h})^2}.
\end{equation}
In order to compute the finite connectivity corrections to $E_{\infty}$, let us expand the cavity field distribution around the limit \eqref{distribCavAsintotica}:
\begin{equation}
    p_z(h)=-\frac{\dd\Phi_z(h)}{\dd h}=p(h)+\frac{p^{(1)}(h)}{z}+o\left(\frac{1}{z}\right),
\end{equation}
from which we have
\begin{multline}\label{Ezexpansion}
 E_z=2\int_0^{+\infty}\dd w\,w\!\iint\dd h_1\dd h_2\theta(h_1+h_2-w) p(h_1)p(h_2)\\
 -\frac{4}{z}\int_0^{+\infty}\dd w\,w^2\!\iint\dd h_1\dd h_2\,\theta(h_1+h_2-w) p(h_1)p(h_2)\\
 +\frac{4}{z}\int_0^{+\infty}\dd w\,w\!\iint\dd h_1\dd h_2\,\theta(h_1+h_2-w) p^{(1)}(h_1)p(h_2)+o\left(\frac{1}{z}\right),
\end{multline}
that, by using the explicit expression of $p(h)$, becomes
\begin{equation}\label{A1}
 E_z
 =\frac{\zeta(2)}{2}-\frac{2\zeta(3)}{z} +\frac{2}{z}\int\dd h\, \Phi^{(1)}(h)\ln\left(1+\e^{2h}\right)+o\left(\frac{1}{z}\right).
\end{equation}
To evaluate $\Phi^{(1)}(x)=\int_x^{+\infty}p^{(1)}(h)\dd h$ we start from its definition
\begin{equation}
    \Phi^{(1)}(x)=-\lim_{\mathclap{z\to+\infty}}z^2\frac{\dd\Phi_z(x)}{\dd z}=-\lim_{\mathclap{z\to+\infty}}z^2\frac{\dd}{\dd z}\left(1-\mediaE{\Phi_z(w-x)}\right)^{z-1}.
\end{equation}
Observe now that
\begin{multline}
\mediaE{\Phi_z(w-x)}=\frac{2}{z}\int_{0}^\infty \e^{-\frac{2w}{z}}\Phi_z(w-x)\dd w=\\
=\frac{\ln(1+\e^{2x})}{z}+\frac{1}{z^2}\left(2\int_0^\infty\Phi^{(1)}(w-x)\dd w+\mathrm{Li}_2\left(-\e^{2x}\right)\right)+o\left(\frac{1}{z^2}\right),
\end{multline}
where we have introduced the polylogarithm of order $s$,
\begin{equation}
    \mathrm{Li}_s(z)\coloneqq\sum_{p=1}^\infty \frac{z^p}{p^s},\quad |z|<1,
\end{equation}
defined over $\mathds{C}$ by analytical continuation. We can write a Volterra integral equation 
\begin{equation}
\Psi(x)+2\int_{-x}^\infty \Phi(t)\Psi(t)\dd t=A(x),
\end{equation}
for the auxiliary function
\begin{equation}
\Psi(x)\coloneqq\frac{\Phi^{(1)}(x)}{\Phi(x)},
\end{equation}
where we have introduced
\begin{equation}
A(x)\coloneqq \ln(1+\e^{2x})-\frac{\ln^2(1+\e^{2x})}{2}-\mathrm{Li}_2\left(-\e^{2x}\right).
\end{equation}
A differential equation for $\Psi$ is immediately written as
\begin{equation}\label{PsiEq}
\Psi'(x)+2\Phi(-x)\Psi(-x)=A'(x),
\end{equation}
or, equivalently, as a system of differential equations for the symmetric and the antisymmetric parts of $\Psi$,
\begin{equation}
    \begin{dcases}
    \Psi_s'(x)=A'_s(x)+\Psi_a(x)-\tanh(x)\Psi_   s(x),\\
    \Psi_a'(x)=A'_a(x)-\Psi_s(x)+\tanh(x)\Psi_a(x).
    \end{dcases}
\end{equation}
Here we have used the fact that $\Phi(x)+\Phi(-x)=1$ and $\Phi(x)-\Phi(-x)=-\tanh(x)$. Moreover, given a generic function $f(x)$, we have used the notation
\begin{equation}
f_s(x)\coloneqq\frac{f(x)+f(-x)}{2},\qquad
f_a(x)\coloneqq\frac{f(x)-f(-x)}{2},
\end{equation}
for its symmetric and antisymmetric part, respectively. After simple manipulations, we obtain
\begin{equation}
\begin{dcases}
\Psi_a''(x)=A_a''(x)-A_s'(x)+\tanh(x)A_a'(x),\\
\Psi_a'(x)=A'_a(x)-\Psi_s(x)+\tanh(x)\Psi_a(x).
\end{dcases}
\end{equation}
We can solve directly for $\Psi_a$ using the first equation. To fix the two integration constants that appear in the integration, we use the fact that $\lim_{x\to\pm \infty}\Phi^{(1)}(x)=0$, and the symmetry conditions (that imply $\Psi_a(0)=0$). We obtain
\begin{equation}
    \Psi_a(x)=-\frac{\zeta(3)}{2}+x+\int_{-\infty}^{x}\ln\left(1+\e^{-2t}\right)\ln\left(1+\e^{2t}\right)\dd t.
\end{equation} From $\Psi_a(x)$ we can directly obtain $\Phi^{(1)}(x)$ as
\begin{equation}\label{Phi1}
    \Phi^{(1)}(x)=\Phi(x)\left[\Psi_a(x)+\Psi_s(x)\right]=\frac{A'_a(x)}{1+\e^{2x}}-\frac{\dd}{\dd x}\left[\Psi_a(x)\Phi(x)\right].
\end{equation}
This expression gives us $p^{(1)}(h)$, see Fig.~\ref{fig:DistribuzCorrStandard}. The explicit expression of $\Psi_a(x)$ is, however, not necessary to calculate the integral in Eq.~\eqref{A1}. Indeed, let us first observe that, due to the fact that $\Phi(-x)=\e^{2x}\Phi(x)$,
\begin{equation}
    \Psi_a(x)\Phi(x)=\frac{\Phi(x)}{2}\left[\frac{\Phi^{(-1)}(x)}{\Phi(x)}-\frac{\Phi^{(-1)}(-x)}{\Phi(-x)}\right]=\frac{\Phi^{(-1)}(x)-\e^{-2x}\Phi^{(-1)}(-x)}{2}.
\end{equation}
Inserting Eq.~\eqref{Phi1} in Eq.~\eqref{A1}, the last integral contains in particular the term
\begin{multline}
    -\int_{-\infty}^\infty\frac{\dd}{\dd x}\left[\Psi_a(x)\Phi(x)\right]\ln\left(1+\e^{2x}\right)\dd x=\int_{-\infty}^{+\infty}\frac{\e^{2x}\Phi^{(1)}(x)-\Phi^{(1)}(-x)}{1+\e^{2x}}\dd x\\
    =\int_{-\infty}^{+\infty}\frac{\Phi^{(1)}(x)}{1+\e^{-2x}}\dd x-\int_{-\infty}^{+\infty}\frac{\Phi^{(1)}(-x)}{1+\e^{2x}}\dd x=0.
\end{multline}
Eq.~\eqref{A1} reduces to
\begin{equation}\label{A2}
 E_z
 =\frac{\zeta(2)}{2}-\frac{2\zeta(3)}{z} +\frac{2}{z}\int\dd h\, A'_a(h)\frac{\ln\left(1+\e^{2h}\right)}{1+\e^{2x}}+o\left(\frac{1}{z}\right).
\end{equation}
Plugging in the equation the explicit expression of $A'_a(x)$, we found that the last integral is equal to $\zeta(3)$. The $\sfrac{1}{z}$ correction in the MP is therefore equal to zero, as numerically verified.

\paragraph{Loopy fractional random matching} The arguments above can be repeated in the case of the LMP. Let
\begin{equation}
    \Phi_z(x)=\int_x^\infty p_z(h)\dd h,\quad \hat\Phi_z(x)=\int_x^\infty \hat p_z(s)\dd s.
\end{equation}
Eqs.~\eqref{cavityeqloopy} imply
\begin{subequations}\label{PhiLoopyEqs}
\begin{align}
\Phi_z(x)&=\left(1-\mediaE{\Phi_z(w-x)}\right )^{z-1}\int_x^\infty\varrho_z(w)\dd w=\e^{-\frac{2x}{z}\theta(x)}\left(1-\mediaE{\Phi_z(w-x)}\right )^{z-1},\\ 
\hat\Phi_z(x)&=\left(1-\mediaE{\Phi_z(w-x)}\right)^{z}.
\end{align}
\end{subequations}
Let us start from the equation for $\Phi_z(h)$. It is evident that, for $z\to+\infty$, $\Phi_z(x)\to\Phi(x)$ as in the MP. Proceeding as in the standard case (and following the same notation) we obtain the same integral equation Eq.~\eqref{PsiEq}, but with a different function $A(x)$, namely
\begin{equation}
    A(x)=\ln(1+\e^{2x})-\frac{\ln^2(1+\e^{2x})}{2}-\mathrm{Li}_2\left(-\e^{2x}\right)-2x\theta(x).
\end{equation}
The next steps follows exactly as in the MP case. We obtain an expression for the antisymmetric part $\Psi_a(x)$ of $\Psi(x)=\frac{\Phi^{(1)}}{\Phi(x)}$ as
\begin{equation}
\Psi_a(x)=\frac{\zeta(2)-2\zeta(3)}{4}+x^2\theta(x)+\frac{1}{2}\mathrm{Li}_2\left(-\e^{2x}\right)+\int_{-\infty}^x\ln\left(1+\e^{-2t}\right)\ln\left(1+\e^{2t}\right)\dd t
\end{equation}
from which the expression of $\Phi^{(1)}$, and then $p^{(1)}(h)$, can be obtained, see Fig.~\ref{fig:DistribuzCorrLoopy}. As before, the $\sfrac{1}{z}$ correction is obtained from $A_a(x)$: from Eq.~\eqref{costoloopycav}, we have that an additional $\sfrac{1}{z}$ correction must be included, due to self-loops and given in Eq.~\eqref{costoselfloops}, obtaining
\begin{multline}
 E_z^{\mathrm{L}}
 =\frac{\zeta(2)}{2}+\frac{\zeta(2)}{2z}-\frac{2\zeta(3)}{z} +\frac{2}{z}\int\dd h\, A'_a(h)\frac{\ln\left(1+\e^{2h}\right)}{1+\e^{2x}}+o\left(\frac{1}{z}\right)\\
  =\frac{\zeta(2)}{2}+\frac{\zeta(2)}{2z}-\frac{2}{z}\int\dd h\, \frac{\ln\left(1+\e^{2h}\right)}{1+\e^{2x}}+o\left(\frac{1}{z}\right)=\frac{\zeta(2)}{2}-\frac{\zeta(2)}{2z}+o\left(\frac{1}{z}\right).
\end{multline}

\section*{Bibliography}
\bibliographystyle{iopart-num}
\bibliography{bibliografia.bib}
\end{document}